\documentclass[aps,prl,preprint,groupedaddress]{revtex4}
\usepackage{graphics,graphicx}
\usepackage{amsmath}

\begin{document}

\title{Electrodynamics in Iron and Steel }
\author{ John Paul Wallace \\ Casting Analysis Corp. }
\medskip

\bigskip
\date{2 June 2009}

\begin{abstract}

In order to calculate the reflected EM fields  at low amplitudes in
iron and steel, more must be understood about the  nature of long
wavelength excitations in these metals.  A bulk piece of iron is a
very complex material with microstructure, a split band structure,
magnetic domains and crystallographic textures that affect domain
orientation.  Probing iron and other bulk ferromagnetic materials
with weak reflected and transmitted inductive low frequency fields
is an easy operation to perform but the responses are difficult to
interpret because of the complexity and variety of the structures
affected by the fields. First starting with a simple single coil
induction measurement and  classical EM calculation to show the
error is grossly under estimating the measured response.  Extending
this experiment to measuring the transmission of the induced fields
allows the extraction of three dispersion curves which define these
internal fields. One dispersion curve yielded an exceedingly small
effective mass of $ 1.8 \times10^{-39} $kg ($ 1.3 \times10^{-9} m_e
$) for those spin waves.  There is a second distinct dispersion
curve more representative of the density function of a zero momentum
bound state rather than  a propagating wave. The third dispersion
curve describes a magneto-elastic coupling to a very long wave
length propagating mode.  These experiments taken together display
the characteristics of a high temperature Bose-Einstein like
condensation that can be initiated by pumping two different states.
A weak time dependent field drives the formation of coupled $J=0$
spin wave pairs with the reduced  effective mass reflecting the
increased size of the coherent state. These field can dominate
induction measurements   well past the Curie temperature.
\emph{Single sensor and high temperature transmission data first
presented as, Crystals for a Spin Wave Amplifier, at 20th Conference
of Crystal growth and Epitaxy, AACGE/West,6 June 2006, Fallen Leaf
Lake, California. }
\end{abstract}

\maketitle

\section{ Introduction}

This review of old data covers some neglected physics  in the
interpretation of low frequency reflection and transmission  of weak
inductive fields in ferromagnetic materials. Historically these
experimental effects are not new and have been detected and reported
as anomalously large permeability measurements of ferromagnetic
material by a time dependent technique introduced by H. Rowland(1)
in 1873 using a pair of inductors coupled by a ferromagnetic torus.
Extending the technique to high temperature measurements of iron in
1910 by E.M. Terry(2) and measurements in high purity iron by R.M.
Borzoth(3) 1937 produced values for permeability orders of magnitude
greater than statically determined values.  The details of the
magnetic domain motion in the ferromagnetic torus were examined in
 1949(4).   The application of Ampere's law assuming no
internal sources of time dependent fields other than the source
inductor was the basis of this transformer design and application.
Here a simplified form of this measurement is considered for
isolating the material contributions to the responses that are not
otherwise described by Ampere's law.

In addition, to my own problem of satisfactorily calibrating
induction measurements on  iron and steel to detect defects there is
a set of other unanswered questions. The storage mechanisms involved
in the operation of ferro resonant load leveling transformer
patented by J. Sola(5) in 1954 require a more detailed explanation.
The anomalous eddy current loss(6)  which is a large effect for
induction heating in Fe and FeSi alloys also has not been accurately
described. Also, the steel corrosion inspection method "remote field
testing" using slowly propagating, highly insensitive long range
fields in low carbon steels(7) also requires an explanation. These
unsolved problems coupled with the large conflict in the simplest
experimental field reflection data to classical calculated values
for ferromagnetic materials exhibits a weakness in our
understanding. Ferromagnetism is a quantum mechanical phenomena, but
a macroscopic connection to the field equations of Maxwell is
essential  to  solving induction measurement problems. Since the
time of Roland's introduction of the coupled transformer experiment,
microscopic theories of elementary spin excitations have been
developed to describe the temperature dependence of magnetization.
The question of how a slowly varying time dependent induction $
\textbf{H}(t) $ interacts with individual spins, spin waves and
structures such as magnetic domain boundaries (MDB) in iron and
steel to produce the measured responses has remained an open
question.

Spin wave theory developed as a lattice theory starting with
considerations of the mechanics of a single one dimensional array of
spins(8,9).   These early works form the basis for the  extensive
investigations that have continued to the present.  One difficulty
in the application of a lattice theory to iron is that it does not
include the local lattice relaxation dependence to the
magnetization. This is a large effect in iron and can be seen in the
fitted plot of lattice parameter as a function of temperature(10)
through the Curie point in figure 1. In this plot the  linear
expansion with temperature is removed so the non linear response of
the lattice to temperature variations taken through the Curie point
become evident.   The strains associated with the material below the
Curie point are significant.

\begin{figure}
  \includegraphics[width=6in]{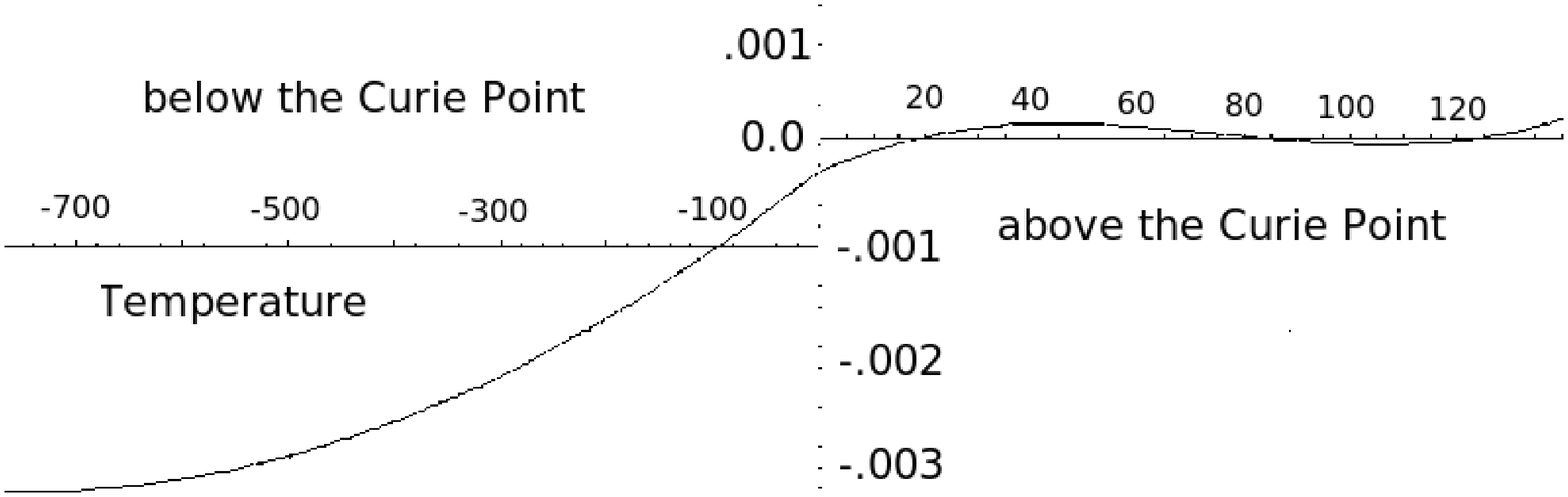}
  \caption{\textbf{ Lattice parameter variations reduced from powder x-ray lattice parameter
  data for iron as a function of temperature with the linear thermal response removed.
    The temperatures are referenced from the Curie point.
  The displacements are in angstroms with the lattice parameter of iron being ~2.86 angstroms at $ 30^o $C.
    The curve is similar to the saturation magnetization as a function of
temperature. The total equivalent
  strain from room temperature to 770 $C^o$ is about $ .1 \% $ due to magnetic ordering. }}\label{Dispersion Curve}
\end{figure}

Spin waves and  photons are the bosons that allow an exchange of
spin states between carriers in ferromagnetic metal such as iron.
These exchanges in addition to the transport of carriers  are how
the spin system is modified by the application of external fields
and induced currents. These two classes of effects in addition to
the lattice relaxation, phonon scattering, determines the measured
responses and the various ways in which energy is distributed or
radiates from the metal. Non saturated soft ferromagnetism that will
be considered are partitioned by an array of MDB and the coupling
between these regions will control the propagation of long range
fields through the bulk of the sample. This couplings can be
summaries in the first table.

\bigskip

\textbf{Table 1:  Field and Current Couplings }
\begin{center}
  \begin{tabular}{|c|c|c| }
  \hline
\textbf{Region} &\textbf{ Coupling } &   \textbf{ Details }\\
\hline
bulk & strain or phonon         & coupling to spin wave in limited freq. range \\
bulk & current                  & $ \textbf{j} \times  \textbf{m}$
spin wave generation \\
bulk & $ \textbf{H}(\omega)$    &  field couples action across MDBs \\
\textbf{MDB}  & current                  &  electron transport across \textbf{MDB}  generate spin waves           \\
\textbf{MDB}  & spin wave                &  emission and capture across \textbf{MDB}  move \textbf{MDB}      \\
\textbf{MDB}  &  photon                   &  emission and capture
across  \textbf{MDB} move \textbf{MDB} \\
Surface & photon & emission and capture  with spin waves  \\
 \hline
  \end{tabular}
\end{center}
\bigskip

\begin{figure}
\includegraphics[width=5in]{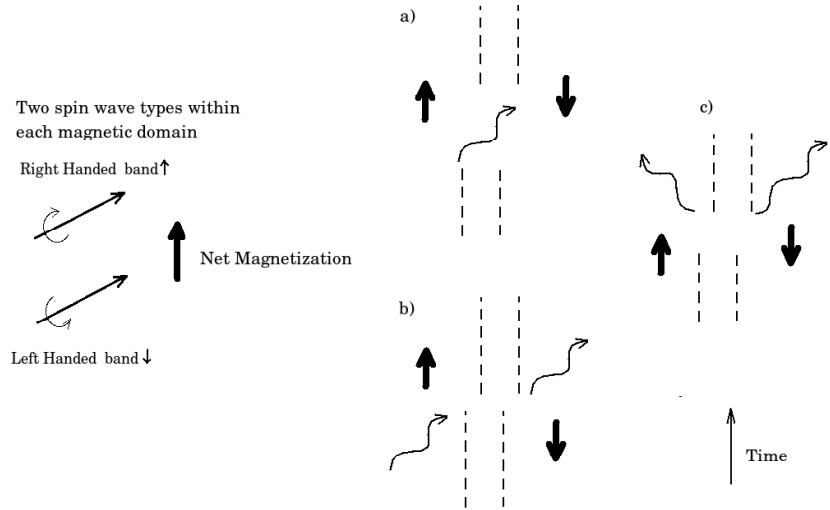}
\caption{ \textbf{ Graphic of the principal transitions that can
generate a non thermal boson population.   The magnetic domain
boundaries are shown as broken parallel lines. For iron which has
two spin bands, there are more paths   available to pump the spin
wave population such as shown in figure 2c which s requires a spin
wave on each band to be emitted.}\label{1020 Hot Rolled Steel}}
\end{figure}

In iron  these 6 process are active at once. Iron's split band
structure(11) allows for direct transitions between electronic spin
states, that are not available in the other ferromagnetic transition
metals and provides another source for non thermal spin waves.  None
of these fields are long range fields within a ferromagnetic metal.

\section{Analysis in Weak Fields}

The construction of the Rowland transformer is  too complex to allow
a simple closed form analysis to represent the experiment. You have
to eliminate the bends to get   a simple experimental  apparatus
with a source inductor around a cylinder and  a receiving inductor
either being the same inductor, figure 3a, or a displaced inductor,
figure 3b. These geometries allows the EM boundary value problem to
be easily solved in closed form using only assumptions of  material
homogeneity.

Our first measurements  are for weak time dependent fields with long
free space wavelengths that will act as a small perturbation on
iron's spin system. For the applied induction field interacting with
a ferromagnetic conductor there are three regions in the material
that influence the interaction. The first is a near surface  where
some domain motion is inhibited by the effect of the surface to
minimize leakage fields. The next deeper region penetrated by the
induction field can drive domain motion more easily.  Finally, there
is the bulk interior below the electromagnetic skin depth  where any
measurable transverse propagating  fields  vanish.

All the measurements described are macroscopic measurements on
physically large samples when compared to grain size, electronic
mean free path and  the conductors electromagnetic skin depths.
These measurements  are on a  small scale when compared to free
space electromagnetic wavelengths and acoustical wavelengths of the
fields at the applied frequency.

\begin{figure}
  \includegraphics[width=6in]{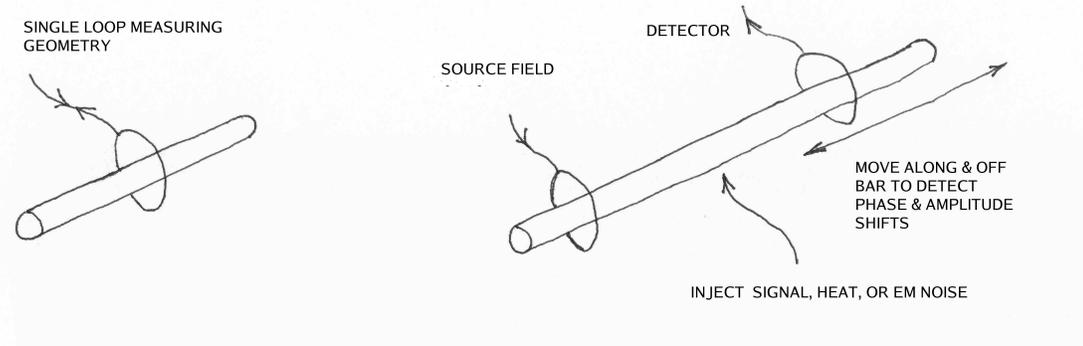}
  \caption{ \textbf{a) Single sensor encircling , b) Dual coil transmission measurement with possible second source injection.}}\label{High Purity Iron}
\end{figure}

The measurements made here will  represent one of a set of
experiments that will isolate the non classical effects  that are
common and easily detectable at low levels with low frequency,
time-dependent fields that interact with ferromagnetic conductors
and insulators.

\section{ Calibration of Induction Measurements}

The principle technique used here is often referred to AC magnetic
susceptibility measurements and  this is a label which is
misleading. For simple conductors the technique is very powerful and
accurate in reducing  measured reflections to structural and
electrical conductivity profiles.   For magnetic material this
technique must assume a magnetic constitutive relation that may not
be accurate or adequate. Therefore, these measurements will simply
be called induction measurements with an analysis based on a
macroscopic set of Maxwell's equations.

An induction measurement system consists of two parts. A network
that supplies and detects the field and the sensor which is usually
a coil. This coupled system is calibrated as a complete system. The
calibration requires the measurement of a standard of know
properties in a computable boundary value geometry that reduces the
standard to a calculable response. The electromagnetic induction
 boundary value problem was  solved
in closed form for selected geometries by Dodd and Deeds(12) in the
low frequency range below 10 MHz for typical metallic conductors. In
the quasi-static approximation for conductivity these solutions are
accurate for simple conductors such as copper, where the electronic
mean free path is short relative to the structure of the sample
being measured(13). This allowed induction measurement of non
ferromagnetic conductors to produce precise dimensional and
conductivity measurements  at ambient(14) and at high temperatures.

What follows is a simple calibration scheme for the fields which can
be though of as an analog to optical reflection. In order, to do
induction studies of high temperature materials, crystal growth
processes a simple calibration technique was introduced(15) where a
known conductivity standard is compared to an unknown sample.
Calibration was reduced to practice when complex arithmetic could be
done on a microprocessor(16) to allow the mapping of a signal space
into a measurement space. This essentially removes the details of
the network that connected the source and receivers so long as they
are operated in a linear response region well below any self
resonance of the measurement system.  This in effect removes the
measurement system by the calibration  in using a two dimensional
transformation matrix for each frequency of measurement. Extending
this calibrated measurement technique to ferromagnetic systems such
a carbon steels, pure iron, cast irons, ferromagnetic ore bodies,
ferromagnetic amorphous metals, ferro fluids and composite
ferromagnetic material produces results that conflict with simple
magnetic models. The assumption required for electrical conduction
analysis is quasi-static approximation and with ferromagnetic
material this assumption is insufficient.  The quasi-static
approximation for electron scattering in conduction does not have an
analogue when considering a system of ferromagnetic spins and their
response to a time dependent field.

For example the simple geometry, figure 3, of measurement and
analysis is a single turn induction loop used both as a source and
receiver surrounding a right circular cylinder of  copper or a well
annealed iron.  Solving the boundary value problem of the wave
equation for the vector potential, $ \textbf{A }$, in a source free
region such as the interior of a conductor will yield the local
electric and magnetic fields through the vector potential.  The
field equation in the source free region where we have used the
definition $ \textbf{B} = \mu \textbf{H} $ is simply:

$$ \nabla^2\textbf{A} - \sigma\mu{\partial \textbf{A} \over \partial t} -
\epsilon\mu{\partial^2 \textbf{A} \over \partial t^2 } = 0 \eqno[1]
$$

The above expression is easily solved by maintaining the continuity
of both the electric field, $ \textbf{E} $ and the magnetic
induction, $\textbf{H} $, at each boundary for either one or two
dimensions with multiple layers.

 The known
properties of the measurement are the signal level, frequencies, the
diameter of the loop inductor and the diameter of the bars under
test.   If a conductor of known purity, dimension and temperature is
used as a standard one can capture the complex signal as reflected
from this standard $ S_m = (x_m,y_m) $ and also compute the
reflected response, $ S_c = ( x_c,y_c )$.  The computed value of the
reflection is found  by solving the boundary value problem. One can
compute a $ 2\times2 $ matrix $ \textbf{T} $, that connects these
two vectors.

$$ S_c = \textbf{T} S_m \eqno[2] $$

A simple example with the  source normalized to $S_{source} =(1,0)
$, then a perfect conductor filling an encircling loop will produce
a response of $ S_c(\sigma\rightarrow\infty) = (-1,0) $ and for an
empty sensor $ S_c =  (0,0) $. For subsequent measurements, $ S_m$,
the application of the matrix $ \textbf{T} $ , transfers the signal
vector into a space that can be interpreted, $R$.

$$ R = \textbf{T}S_m  \eqno[3] $$

  It is useful to consider the limits of these
measurements and responses in a simple experiment.   With fixed coil
or loop surrounding a cylindrical conductor the response $
R(\sigma)$ where $ \sigma $ is the conductivity will obey this
relationship,

$$ K = {|R(\sigma)| \over |R(\sigma\rightarrow\infty)| } < 1 \eqno[4] $$

This can be shown by solving the boundary value problem and
observing the behavior of the responses by taking the cylinder's
conductivity to infinity.  This is simple to show either for a long
encircling coil about a cylinder, a plane wave on an infinite plane
reflector or a single loop surrounding a cylinder. Similarly if the
material permeability is examined in the same way the result are the
same. There can be no measured reflected signal that was greater
than initially supplied. This is easy to prove in any geometry. This
is because the propagation vector in the material contains a product
of $ \sigma $ and $ \mu $ and even if complex the limiting relations
of equation 4 and 5 remain true.

$$ \frac{|R(\sigma ,  \mu)| }{|R(\sigma \rightarrow\infty ,  \mu_o)| }  \leq  1\eqno[5]  $$

An example of this type of computation for a simple one dimensional
reflection deriving the upper bound on K is located in appendix A.
At low field levels  equation 4 can be understood as  a statement of
the dissipation as a result of material resistivity. The denominator
is the response of a very good conductor. The numerator describes a
similar sample where we are increasing the permeability possibly by
increasing the sample temperature. Equation 5, however, is only a
result of solving Maxwell's equation using the simple linear
magnetic equation of state,

$$  \textbf{B} = \mu \textbf{H}  \eqno[6] $$

The weakness in this model description of a ferromagnetic material
is apparent when the ratio $ K $ is measured at values greater than
1.   This solution is analogous to any optical reflection where the
reflected  amplitude is always less than or equal to the incident
source.

In a good conductor the one dimensional cylindrical solutions are
quite accurate for a long sensor because the fall off in  fields at
the coil ends are abrupt,  $ ~ \frac{1}{r^4} $. Where $ r $ is the
distance from the end of the coil. This rapid decrease results from
the cancelation of the fields due to the out of phase induced field
in the conductor. So standards are typically taken with high
conductivity non magnetic materials.

\section{Measurements}

The effects  to be examined here are the reflection of an imposed
time varying field on various ferromagnetic materials. In addition
the transmission of a signal generated  from a time varying field
will also be examined. Particularly, the penetration of signal well
beyond the electromagnetic skin depth in soft ferromagnetic
conductors. The analysis is  strictly macroscopic with the aim of
determining the changes required in the application of Maxwell's
equation in order to physically understand the responses.  The
equipment  used evolved from early multi frequency eddy current
instruments used on stainless weld measurements(17) where the
response of  minority delta ferrite  and solidifying cast irons(18)
were monitored for phase and magnetic transformations. The signal
generation and detections system used in the current measurements is
a Process Monitor IV from Casting Analysis Corp. which has three
quadrature phase detectors that can provide three signal sources
that  operate independently. All detectors use a common master clock
which maintains a uniform phase relation in time among all channels.
The system can feed all three signals into either a single or
multiple coils that can be detected on any three channels.  The
operating frequency range is from 1 Hz to 20 MHz.  Typically this
instrument is used for multi frequency direct inversion of
calibrated eddy current data for monitoring properties over a wide
temperature range such as crystal growth.  The free channel is used
for spectrum scanning assuring that the system is operating in
linear range where there is not a measurable $ 2f $ or $ 3f$
harmonic from the sample.

The operating software can compute the boundary value problem in the
 geometries that are used in the following experiments as a long
solenoid and a single loop. The quadrature responses are computed
for the standard and then data taken on that standard is used to
compute the matrix $ \textbf{ T}_i $ for each frequency, $ \omega_i
$ of measurement. This allows single or multiple frequency
reflection measurements to be made along with transmission
measurements. When taking transmission measurements the calibration
uses the sample under test as the standard or no sample with the
source and receiver as close as possible. This results in  a
reference at a phase angle of $ 0^o $ so that measurements on the
bar produce the total phase shift as referenced to $ 0^o $ or the
source. The typical maximum drive are 10-20 milliampere for the
reflection experiments into 12 turn coils 1.5 cm in diameter and 1.5
cm long. These coils are driven in series with a 51 ohm resistor.
The sample diameter determines the local maximum for the field. All
detector coils are terminated with a 51 ohm resistor to ground. The
detector coils are the same construction as the probe drive coils.
The current levels are set at least an order of magnitude below
where we cannot detect non linear components in high purity well
annealed iron  when two different frequencies are used to drive the
source. The linear behavior of the system  acting on copper is
confirmed experimentally in data taking with increasing current
inputs showing a well behaved response  as shown in figure 4. The
behavior for iron is different with a response increasing
monotonically with the drive level in the same figure.

\subsection{Simple Cylindrical Boundary Value Reflection Experiment and Analysis}

The first experiment is a simple reflection measurement in the
cylindrical geometry for a set of standard materials when the source
and sensor are the same, show the non classical behavior of the
detected  amplitude to the applied signal.   For a magnetic material
with $ \mu > 1 $  the value of $ \textbf{K} $ should be less than 1,
however, the two magnetic materials measured show values between 22
and 35.    This is a very large error, greater than $ 2000\% $ for
the application of a simple constitutive relation, $\textbf{B} = \mu
\textbf{H}$ .

\medskip

\begin{figure}
  \includegraphics[width=6in]{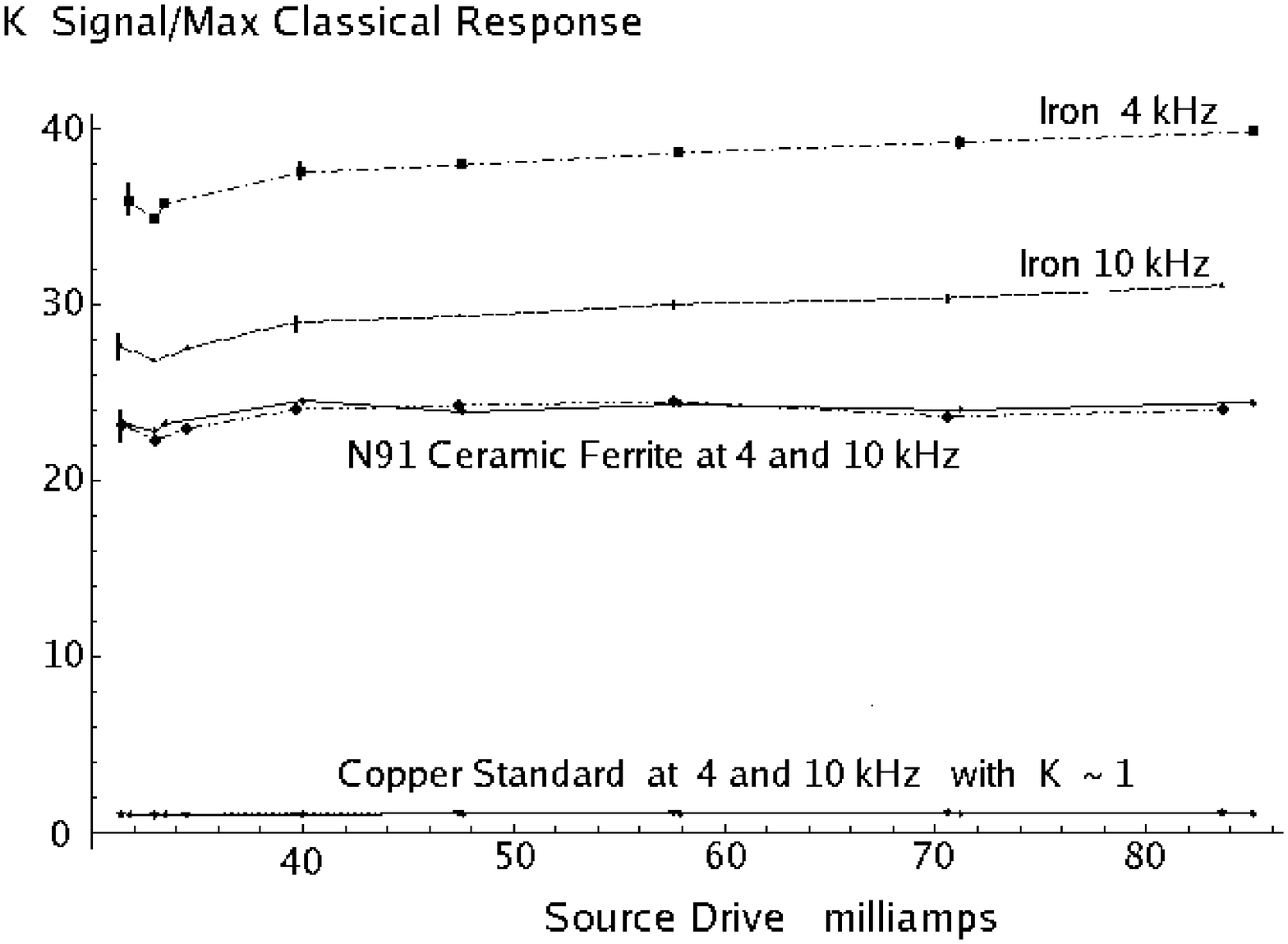}
  \caption{ \textbf{Single sensor measurements of reflected eddy current response in copper,
  iron and a NiZn ferrite. The key feature is that the ferromagnetic
  samples all have K $ \gg $ 1.   In iron's  case  this value approaches 40.  The iron in this case
  was of 99.99$\% $ purity and held for greater than 30 days at 1050 $ C^o $ in flowing hydrogen and then furnace
  cooled to reduce the interstitial content.  Iron differs from the other samples in that there is
  a signal increase as a function of drive current which is not the case for copper or the ferrite.  This nonlinearity is
  also not predicted by the classical modeling in the boundary value problem. The variations at low current levels are caused
  by low signal level for the copper standard resulting in reduced precision. In this experiment field levels range up from
  $ 10^{-7}  $ Tesla. }}\label{High Purity Iron}
\end{figure}

The simplest reflection experimental data is recorded in figure 4
which shows  the signal for  3 samples at two different frequencies
while the source drive level is increased.     The key point to take
from this data is that for soft ferromagnetic conductors and
insulators there is measured a significant signal energy above the
applied level. As shown previously, it is not possible to get a
result found in figures 3 or 4 by varying $ \mu $ in the complex
plane. There is also no apparent resonances detected in the
frequency range as shown in figure 4.

\begin{figure}
  \includegraphics[width=6in]{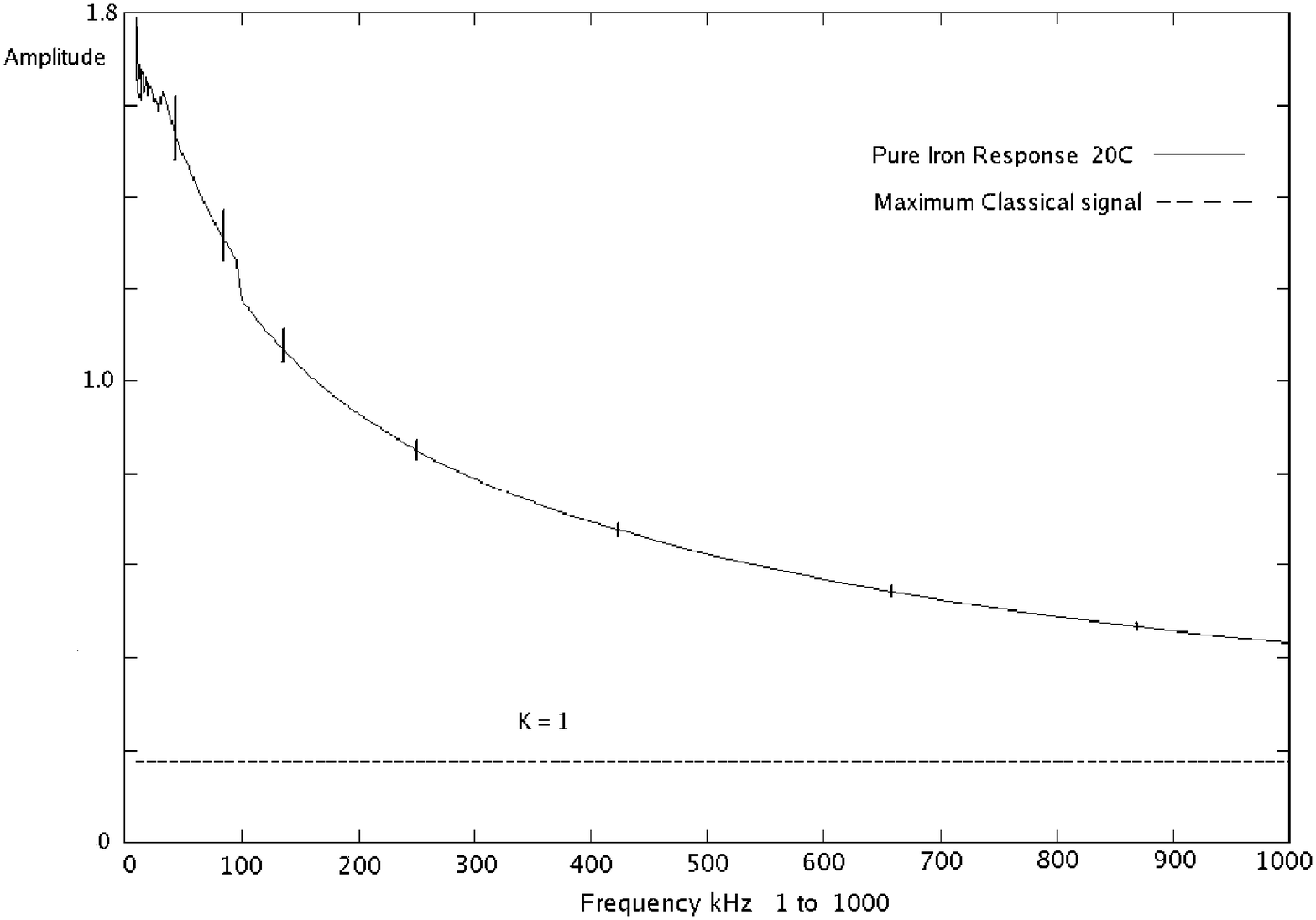}
  \caption{\textbf{Single sensor measurements of  amplitude plotted as a function of frequency in high purity iron.
    Three separately calibrated sweeps are joined to cover the frequency
   range, and the offsets observed are from joining data together.
   The noise at low frequencies is due to less  resolution for the  calibration measurement  on
   copper which is used to normalize the data to
   assembling the sweep on one scale.   }  }\label{High Purity Iron}
\end{figure}

The apparent enhanced level measured  at the applied frequency is
not a peaked resonance as one would find in a  oscillator associated
with a local physical property; because if the frequency is
decreased, there will be a slow increase in the amplitude. By using
$\textbf{B} = \mu \textbf{H}$ there is not a way to compute a return
amplitude greater than 1.  Since ferromagnetism is a quantum
mechanical phenomenon there is not any reason to assume that the
macroscopic field should be proportional to the applied induction.
These large signals also indicate we do not have accurate knowledge
of the spatial distribution of the fields within the cylinder.

\subsection{  Radial Scale Independent Measurement of Long Solenoidal
 Reflection }

 To gain more understanding of the difference of a ferromagnetic
 material such as iron to a simple conductor such as copper in the response,
 it is easy to remove the radial scale dependence from the problem
 of figure 3a to get a material response that is independent of the radius of the
 sample for a known applied field.   If the material behaves as a
 simple reflector with Joule dissipation of the induced currents the reflection
 now should be independent of the sample's radius.   The boundary
 value problem for the continuity of the electric field, \textbf{E}  and the
 magnetic induction \textbf{H} result in a pair of equation at the boundary of
a homogenous cylinder radius, r, where $ k_o $ is the free space
propagation vector.

 $$    A J_1(k_or) +  B Y_1(k_or)  =    C I_1(kr) \eqno[7] $$

 $$   \frac{1}{\mu_o} (A {J'}_1(k_or) + B{Y'}_1(k_or )) =
 \frac{C}{\mu} {{I'}_1(kr)}  \eqno[8] $$

 If the source coil has a radius, $ r_c $,  with a source field  $ A
 J_1(k_or) $ we can consider the field at the surface of the
 cylinder is normalized by the ratio   $ J_1(k_or)/J_1(k_or_c) $
 and similarly  the response from the cylinder  is normalized by the
 ratio $ Y_1(k_or_c)/ Y_1(k_or) $.   The measured responses, S,
 amplitude is then found in the scale independent form, $ S_n $ as:

 $$  S_n =  S \frac{J_1(k_or_c)Y_1(k_or)}{J_1(k_or)Y_1(k_or_c)}  \eqno[9] $$

 Taking data on some annealed low carbon steel and copper of
 different radii  at 10 kHz illustrates the difference in the
 materials' responses.

\bigskip
\textbf{ Table 2:  Scale Independent Response with A Long Solenoid
for Copper and hot rolled 1018 steel, 10 kHz coil 7.85mm Radius.
The raw measured signals in Copper fall fast as the radius is
reduced. The radially independent signal for copper is flat. The
signals actually grow in the hot rolled 1020 steel as the radius
reduces.}

\bigskip

\begin{center}
  \begin{tabular}{|c|c|c|c|c|c|}
  \hline
Cu Radius mm &  S $\pm $ .007 & $ S_n $ & Fe Radius mm & S $\pm $ .007 & $ S_n $ \\
  \hline

 4.75 &  .319 & \textbf{ .87} & 5.85 & 2.00 &\textbf{ 3.60} \\
 4.25 &  .263 &  \textbf{.89}  & 4.85 & 1.54 &\textbf{ 4.03}\\
 1.55 &  .027 &  \textbf{.7 } & 3.95 & 1.276 & \textbf{5.04}\\
      &       &        & 2.65 & .888 & \textbf{7.79 } \\
\hline

  \end{tabular}
\end{center}
\bigskip

The main feature of table 2 is that copper's response is nearly
independent of radius  and the signal for the iron data is large and
 growing with decreasing radius.  The result is an increasing signal with reduction of radius
where the ohmic loss effects have already been corrected as a
function of radius.  As the radius decreases, the path for losses
out the ends is reduced.

The practical coupling of fields to ferrous cores in transformer
design illustrates some advantages for weak time dependent field
when increasing the surface area relative to volume.  It is not
apparent that the traditional explanation of eddy current losses
play much of a role in this optimization for more area per unit
volume in a ferrous core for weak fields.

\subsection{ Saturation Effects on Transmission  and Local Signal
Level }

One simple experiment that can be performed in measuring the locally
reflected field from a cylindrical sample is to apply an axial
static magnetic field of a few hundred gauss from a ceramic
permanent magnetic with a central hole in which a sample can be
placed with a coaxial coil, figure 3a. This has the main effect of
removing a significant fraction of MDB and aligning the static
magnetization within the material.  The effect on the local signal
shows a large difference in the soft magnetic materials. The locally
measured signal and the responses with and without a static axial
magnetic field are listed below.
\bigskip
\textbf{Table 3:
 Static magnetic field  effect on the reflection
for  50 kHz  signal. The error bars on phase  $\pm .02^o $ and
amplitude  $ \pm .002 $.  }
\begin{center}
  \begin{tabular}{|c|c|c|c|}
  \hline
 \textbf{Material}  & \textbf{ phase} $ \phi $  &  \textbf{ change in phase}& $ \% $ \textbf{amplitude reduction} \\
  \hline
Fe       & $ -39.7^o $    &    $   -35^o $    &  $87.5\%$ \\
Ni       & $ -84^o   $    &    $   -63^o $    &  $30\%$   \\
Co       & $ -81^o   $    &    $   -29^o $    &  $27.5\%$ \\
Ferrite N91 & $  -2.7^o $  &   $   +1.5^o $     &  $96.2\%$
\\

\hline
  \end{tabular}
\end{center}
\bigskip

The phase data shows the metals Ni and Co with the saturating fields
decrease with partial saturation. The Ni and Co are from machined
bar stock and not annealed whereas the Fe sample is a well annealed
specimen with very low interstitial impurity content. For a simple
conductor the phase angles fall between $ -90^o
> \phi > -180^o $ whereas for a ferromagnetic insulator like N91 the phase angle will
approach $ 0^o $ or $ S_m = (amplitude,0) $ for a response vector.
The static \textbf{B} field is not quite strong enough to push the
phase angle response of Fe to values as low as $ -90^o $. The NiZn,
N91, ferrite is the softest of the materials and whose amplitude is
greatly effected, but its phase is driven towards $ 0^o $ which is
indicative of a more highly permeable insulating material phase
response for a classically describe magnetic material.   The iron
and the ferrite amplitude response are well beyond what one would
calculate as the limiting response for a classical reflection
without the static field.  With the static field the phase angles
fall into better agreement with a classical analysis. The result of
minimization of MDB area or immobilizing MDB on microstructural
features from cold work acts similarly by removing a significant
portion of the nonclassical response.

\section{Propagating Weak Field  Measurements}

For propagating fields our first measurements  are for weak time
varying fields with long free space wavelengths that will act as a
small perturbation on the iron.   We are within a small fraction of
the  free space wavelength at the source field for the entire
frequency range  being examined.  The dispersion of the components
that produce the measured excesses in the reflection responses may
allow the individual sources and propagation modes to be isolated
and identified. By scanning the displacement of the receiver as well
as the frequency of the source, the overlapping effects that occur
with a single driven sensor/receiver can be partially lifted.

Calibrating the transmission standards is much simpler because it
can be a differential measurement from the position of closest
approach of the source and receiver normalized to a signal vector
 $ S =(1,0)$. The first component representing the magnitude of the
component in phase with the source and the second component is the
amplitude of the signal that  is phase shifted $ -90^o $ to the
source. The zero point is established by removing the drive signal
from the source. This allows the differential phase change to be
acquired accurately with displacement. An absolute calibration can
be made by removing the medium and measuring the source field
directly as is done in the single sensor measurement. The amplitude
responses are scaled to the normalized source amplitude  at the
closest approach. This  allows a direct measurement  of the fields
decay as a function of displacement.

In the transmission experiments, figure 3b, the coil drives are
typically 30 milliampere into 10 turn coil 1.3 cm in diameter and .9
cm long, driven through a 10 ohm series resistor. The sample
diameter determines the local maximum for the field which can be
easily computed. The detector coil is the same as the source coil
and is terminated with a 51 ohm resistor to ground and it feeds a
 transformer input on the Process Monitor IV.  In translation
measurements for the extraction of the dispersion curves Process
Monitor IV steps the probe down the bar taking readings every .05 mm
while running at a single frequency for each scan.  When selecting a
frequency to work at in the transmission mode a frequency sweep
should be made with the source off to determine if there are any
extraneous signal sources being sensed.

\subsection{First Transmission Measurement}

From the literature it is known that induction formed  propagating
fields in iron and steels exist(7). Also we have a great deal of
data(19) on the high temperature measurements of steels with a
single probe which shows a significant enhancement of the field
levels above what we measure at room temperature. An example of this
is shown in figure 6.  The other features of the high temperature
data are the constant phase value  as one approaches the Curie point
and the decreasing noise level detected in the large reflected
response.  In the example in figure 6 the signal increase is by a
factor or 6 moving to the Curie point.  The phase drops at the Curie
point to a value expected of a conductor at above $ 770^o$C.

\begin{figure}
  \includegraphics[width=6in]{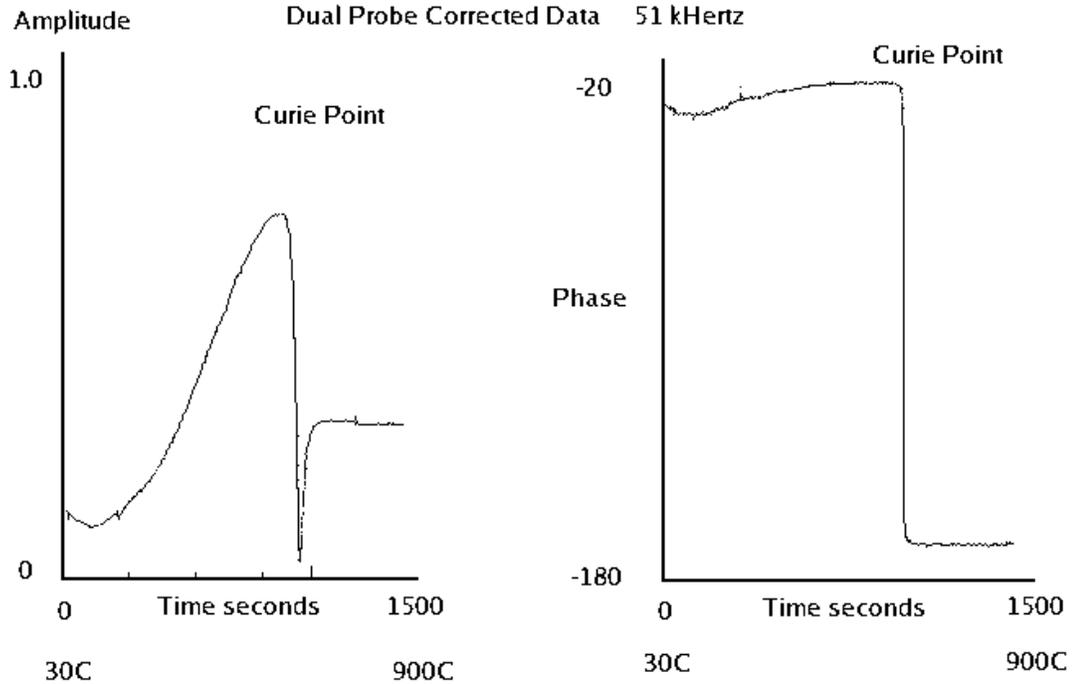}
  \caption{ \textbf{Single sensor on iron at 51 kHz, heated  from 30 C to $ > $ 900 C  reflected amplitude
  response on the $ 0^o $ axis.
  Data taken on a Process
  Monitor II in a low carbon steel.  }\label{High Purity Iron}}
\end{figure}

The question posed here is  whether transmission will be possible in
a region where  the steel is raised above the Curie point? This
transmission experiment depends on two processes, first is
generating the field and the second is the role the medium plays in
its motion.  This immediately would reveal whether the signal could
survive in a paramagnetic region of no permanent magnetic moment but
a spin wave population and no magnetic domain structure. If the
field traversed that region then it answered one question about the
necessary medium required for transmission of the signal. This first
experiment was done on a 4.7 mm diameter rod 1018 hot rolled steel.
The coils were 11 turns mounted on water cooled copper mono turn
rings and 8 mm long. The calibration was based on the source and
receiver coils empty and far removed for a zero point.  Then the
coils brought close together axially for taking the $ 0^o$ phase and
amplitude reference. So that all subsequent measurements are
referred to these measurements and the phase delays are absolute and
caused by the material.

\begin{figure}
  \includegraphics[width=6in]{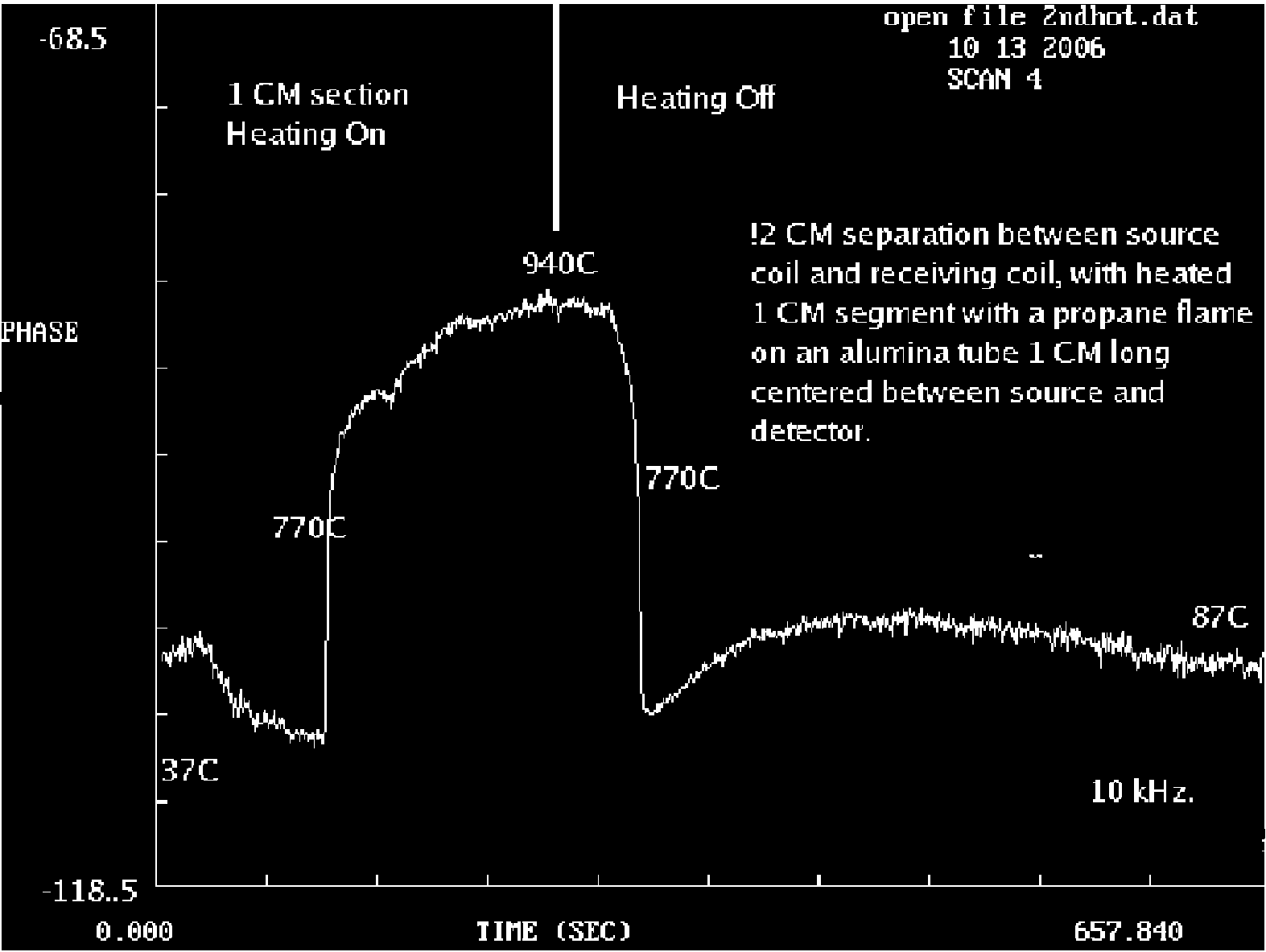}
  \caption{\textbf{ The phase response of the 10 kHz transmission
 over 12 cm through a 1 cm hot zone
  in a  1018 steel rod. The sample starts at $37^o$C and is heated by flame to $940^o$C and allowed
   to cool to $87^o$C. The geometry is similar to figure 3b.} }\label{High Purity Iron}
\end{figure}

\begin{figure}
  \includegraphics[width=6in]{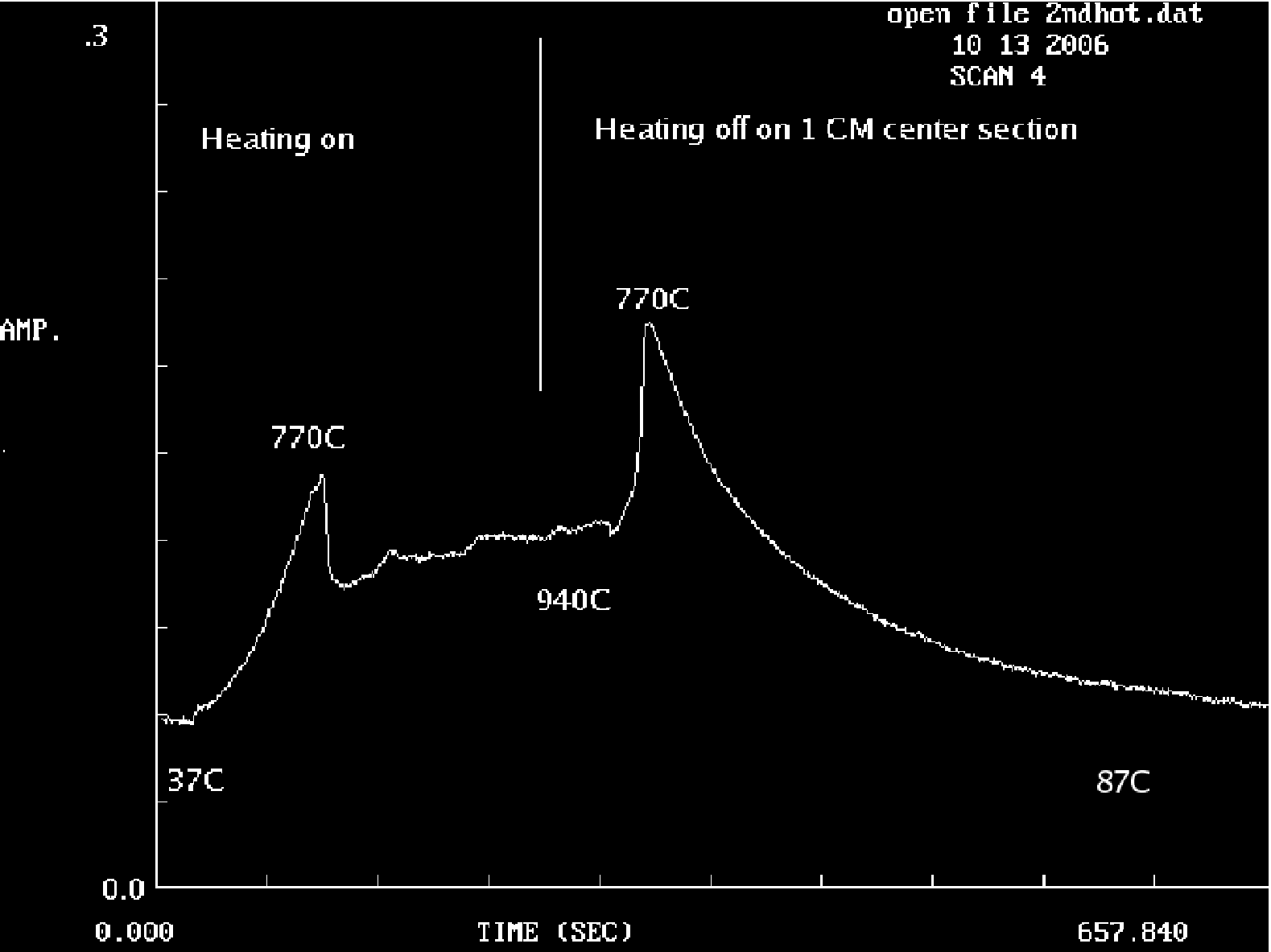}
  \caption{\textbf{The amplitude response of the 10 kHz transmission
    over 12 cm
   through a 1 cm hot zone in a  1018 steel rod. The sample starts at $ 37^o $C and is
   heated by flame to $940^o$C and allowed to cool $ 87^o$C. The geometry is similar to  figure 3b
   and the source and receiver coils were cooled. The asymmetry in the
   response as the flame is removed  represents the cumulative heating of the rod as
   heat diffuses outward from the center changing the gain characteristics.}}\label{High Purity Iron}
\end{figure}

\bigskip

\bigskip

The transmission data for phase and amplitude with the heated center
are shown in figure 7 and 8 respectively. In the vicinity of the
Curie point there was a rapid increase in the phase delay along with
a steep increase in response. Then as the center region passes
through the Curie point the transmission level abruptly drops, not
to zero, but to value that stills shows a transmission response
double that of the room temperature value. This response is made up
of a combination of the gain in the heated rod below the Curie point
and the loss  in the region above the Curie point.  The phase delay
suffers a significant reduction with the section above the Curie
temperatures.  The transition through the Curie point arrests a
portion of the propagating field but from a classical EM analysis
the attenuation is too weak and the phase response is of the wrong
sign.  That is if an axial induction at 10 kHz is impressed on a
conductor it is rapidly attenuated on the order of 1 mm and is phase
is significantly delayed and not advanced.   Where as the single
coil heating experiment of figure 6 shows nothing unusual and
differing from a classical EM response above the Curie point.
Generation of the excess signals seen below the Curie point both in
the transmission and single sensor data are not explained by a
classical analysis.

The frequency dependence of the transmission drop at the Curie point
is shown in table below for a set data take  with the identical set
up as just described but with a transmission distance of 14.5 cm and
a heated zone of 2 cm.

\bigskip

\begin{center}
\textbf{Table 4: Signal transmission through at 2 cm long hot zone
with a spacing between transmitting coil and receiver of  14.5 cm.
The phase decrease error bars, are approximately 10 $\% $ of the
value.}

  \begin{tabular}{|c|c|c|c|c|}
  \hline
 \textbf{  kHz } & \textbf{ Phase Decrease } & $ \% $ \textbf{ Field Blocked}& \textbf{Gain to Curie Point} & \textbf{Comment} \\
  \hline
3    & $ -24^o $ & $ 62 \% $ &  $ 131 \% $ & \\
5    & $ -21^o $ & $ 43 \% $ &  $ 153 \% $ & \\
10   & $ -31^o $ & $ 38 \% $ &  $ 271 \% $ & \\
20   & $ -18^o $ & $ 16 \% $ &  $  85 \% $ & \\
30   & $ -8^o  $ & $ 9  \% $ &  $  38 \% $ & \\
50   & $ -5^o  $ & $ 4  \% $ &  $  48 \% $ & \\
100  & $ -4^o  $ & $ 3  \% $ &  $   1 \% $ & \\
200  & $ -1.5^o $& $ >1 \% $ &  $  >1 \% $ & \\
500  & $ -1^o $  & $ >1 \% $ &  $  >1 \% $ & \\
1000 & $ -.3^o $ & $ >1 \% $ &  $  21 \% $ & on cooling\\
1850 & $ -.7^o $ & $ 1  \% $ &  $   6 \% $ & on cooling \\
2900 & $ -.8^o $ & $ 2  \% $ &  $   3 \% $ & on cooling \\

\\

\hline
  \end{tabular}
\end{center}
\bigskip

 The heating for this particular data was done with a
propane flame and not an electric furnace.   If an electric furnace
is used the results are much more complex depending  on the content
of ripple in a  DC supply  or an AC power source is used because
this induces a  nonlinear interactions with the time varying heating
field from the heating elements.  If you were to replace the heat
source with a local transverse strong static magnetic field at the
center of the bar little change is measured.

\subsection{ Transmission and Dispersion Curves }

\begin{figure}
  \includegraphics[width=6in]{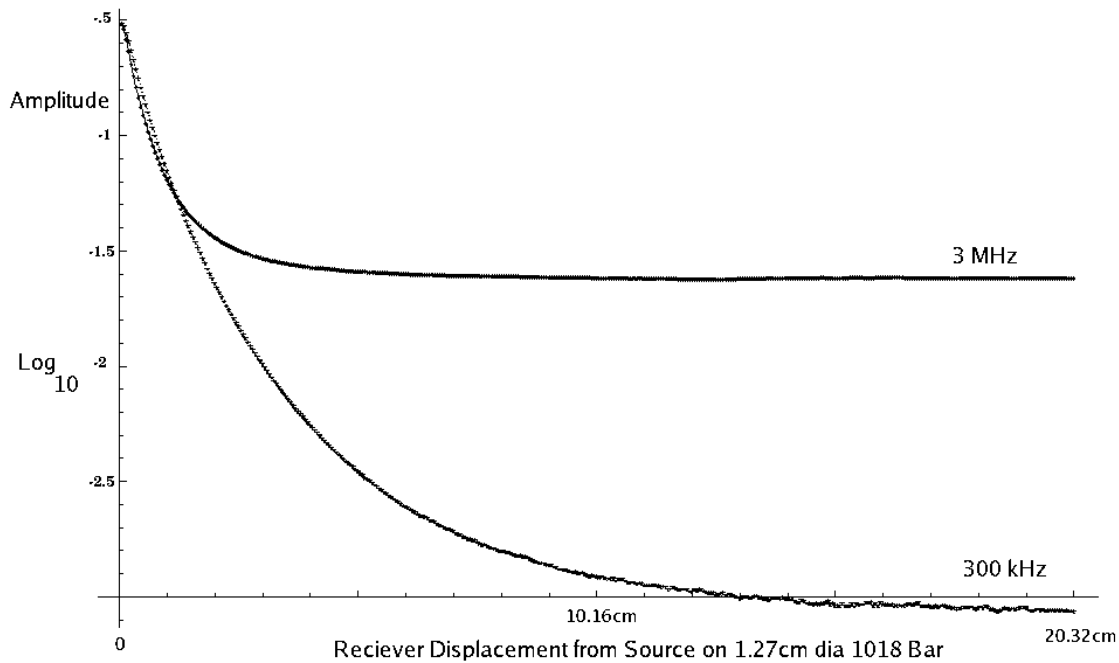}
  \caption{ \textbf{ $ Log_{10} $ Amplitude of a 300 kHz and 3 MHz signal verses separation from source.
  Displacement amplitude sample data above is one of a set of 15 scans at  frequencies from 3 kHz to 3
  MHz.   The data was used to construct the transmission response graph in figure 11 along with the
  phase data for the same scan shown in figure 10. }\label{1020 Hot Rolled Steel}}
\end{figure}

\begin{figure}
  \includegraphics[width=6in]{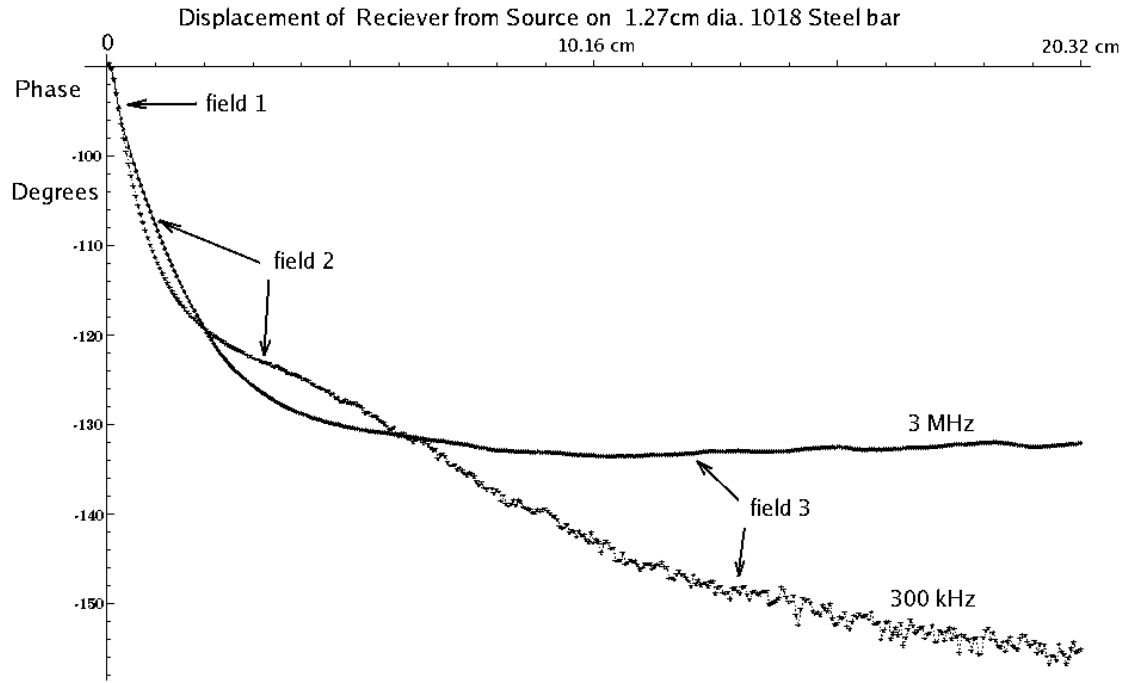}
  \caption{ \textbf{  Phase of a 300 kHz and 3 MHz signal verses separation from source in degrees.
   This is one of a set of scans  from which  phase data and displacement information was gathered.
   Data from these phase and amplitude scans were used to construct the three dispersion curves in figure 11.}\label{1020 Hot Rolled Steel}}
\end{figure}

\begin{figure}
  \includegraphics[width=6in]{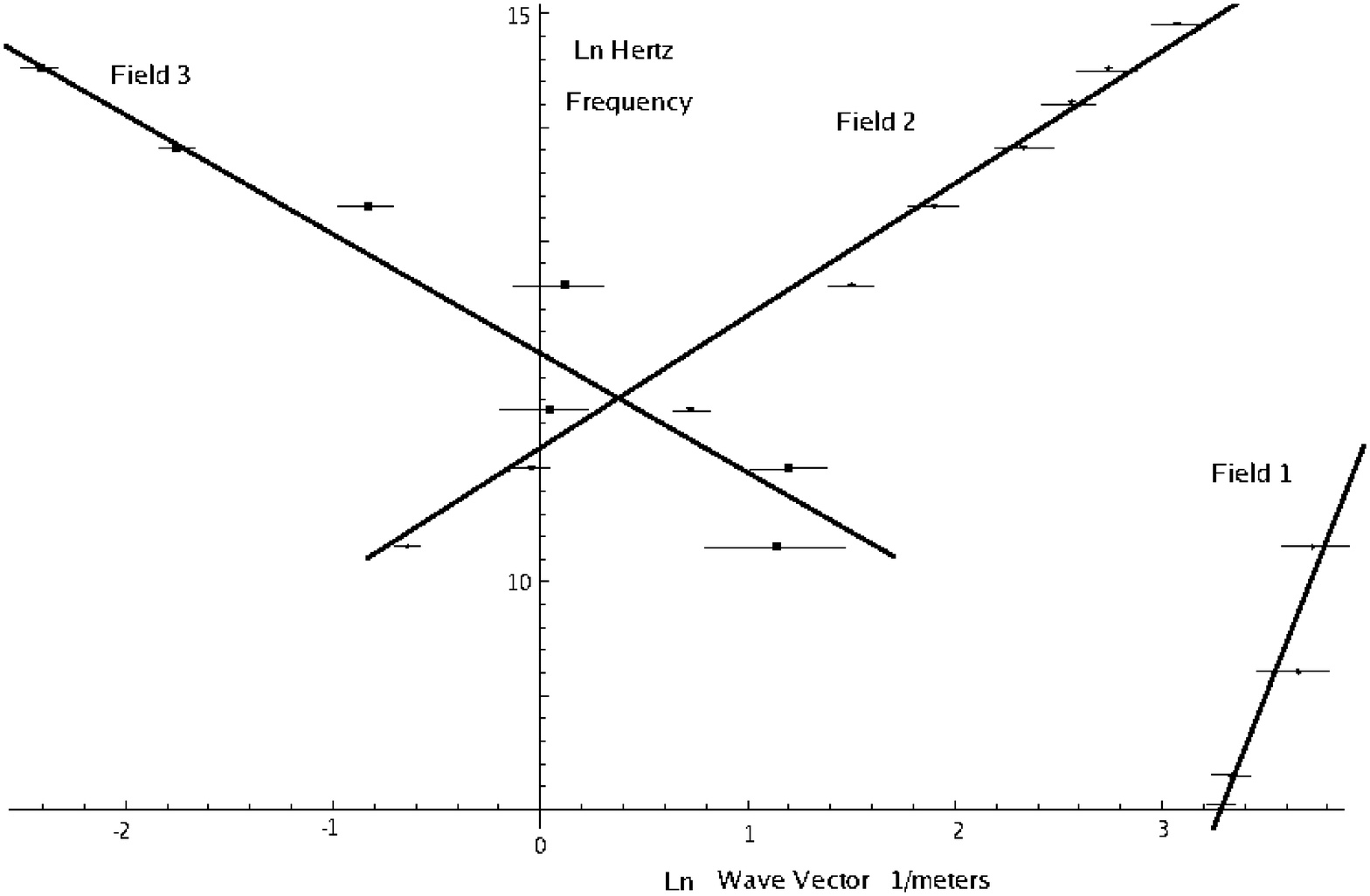}
  \caption{ \textbf{ To fit the three dispersion curves on a single
  graph a  natural Ln-Ln plot in frequency for the vertical scale
  and propagation vector K in 1/meters.  Field 1 has a velocity
  close to that of
  sound and is only seen below 30 kHz. Since the sensor is only sensitive
  to a time dependent magnetization, there would have to be coupling between
  fields.   Field 2 has  a computed
  effective mass of $1.8\times 10^{-39}$ kg ( $1.3\times 10^{-9} m_e $).
  Field  3 dominates above 500 kHz and rolls off very slowly. The
  intersection point of fields 2 and 3 provide for a degeneracy
  point where we could expect properties to be effected strongly.
  }\label{1020 Hot Rolled Steel}}
\end{figure}

The transmission measurement is a very useful experiment because if
a resolvable field as a function of the source frequency can be
identified with a measurable phase-distance relationship then a
dispersion curve for that field can be constructed.  Specifically
this means that both the phase and amplitude data must be well
behaved over a frequency range so that a field can be identified and
enough points taken to reconstruct a dispersion curve. These
measurements were taken in a 1018 hot rolled 12.7 mm diameter steel
rod 69 cm long with the measurements done at $20^o$C. This material
was selected because it is inexpensive and essentially well annealed
iron with a low concentration of cementite precipitates, very little
carbon in solution and  annealed by the slow loss of residual heat
stored in the large coil formed from a single billet after hot
rolling. It suffers only a mild straightening operation with its $
w\ddot{u}stite $ patina in tack. Because of the heating and rolling
in air and the visible oxidation there will be a surface
decarburization band which will leave a relatively pure iron just
below the oxide.

From the linear displacement of the receiving probe two well
resolved fields were observer and one poorly resolved field with
sample raw data from 2 of the 15 scan taken shown in figures 9 and
10. Field 1 below 30 kHz is possible an acoustical field driving a
time dependent magnetization. This field is heavily damped with a
detectable  a range of only 4 cm and estimated velocity of 4
$\times10^3$ m/s is shown in figure 11. This field 1 is only
resolved at the lower frequencies and it could not be resolved above
30 kHz.  The amplitude was difficult to separate from the source
signal but it was significantly greater than 10 $ \%$.

\begin{figure}
  \includegraphics[width=6in]{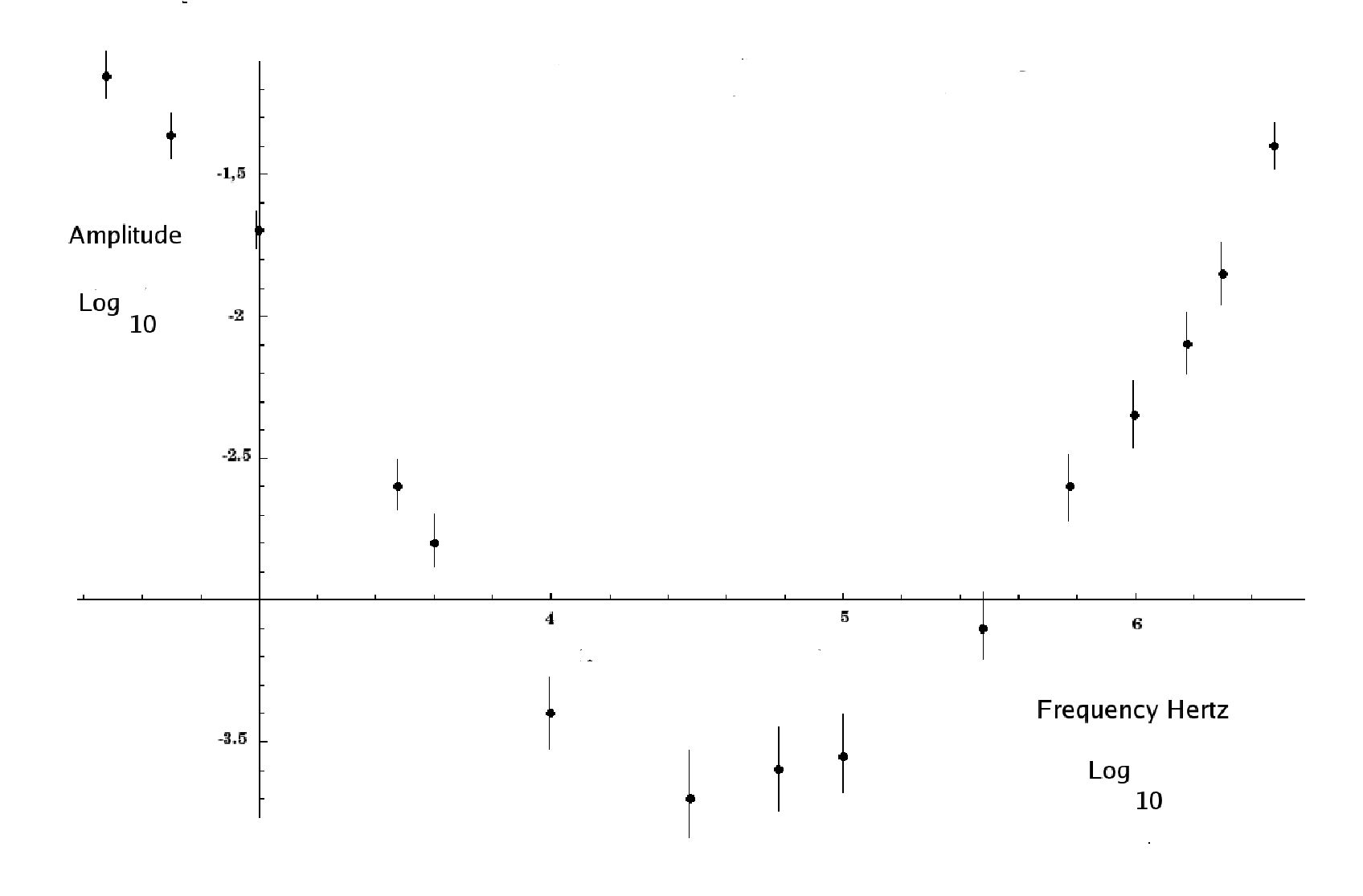}
  \caption{ \textbf{ $ Log_{10} $ Amplitude verse $ Log_{10} $ frequency for long range field at 20.32 cm from source.
  This data has a practical application in that for induction heating one would want to minimize the propagated energy
  loses and that would have a minimum for this particular steel between 10 and 30 kHz.   This minimum is where
  induction heating of steel is found to be most efficient below the Curie point.  This region also corresponds to
  the intersection of field 2 and 3.   This degeneracy could lead to stronger scattering
  between these two states in addition to magneto-acoustic losses in this frequency window.
  }\label{1020 Hot Rolled Steel}}
\end{figure}

The next  dispersion curve is field  2  and it  emerges resolvable
from field 1 at 30 kHz. This field is detectable to 2 MHz, figures
11,13 and 14. This fields frequency has a parabolic dependence on
the propagation vector consistent with the field carrier having a
mass and behaving as a free particle.  This field was not expected
in a room temperature poly crystalline steel.   This very low mass
for field 2   makes it a candidate for a new type of quasi particle
or a highly modified spin wave whose effective mass has been
reduced. Above 2 MHz this field is not resolvable because its decay
closely overlaps the source signal.

\begin{figure}
  \includegraphics[width=6in]{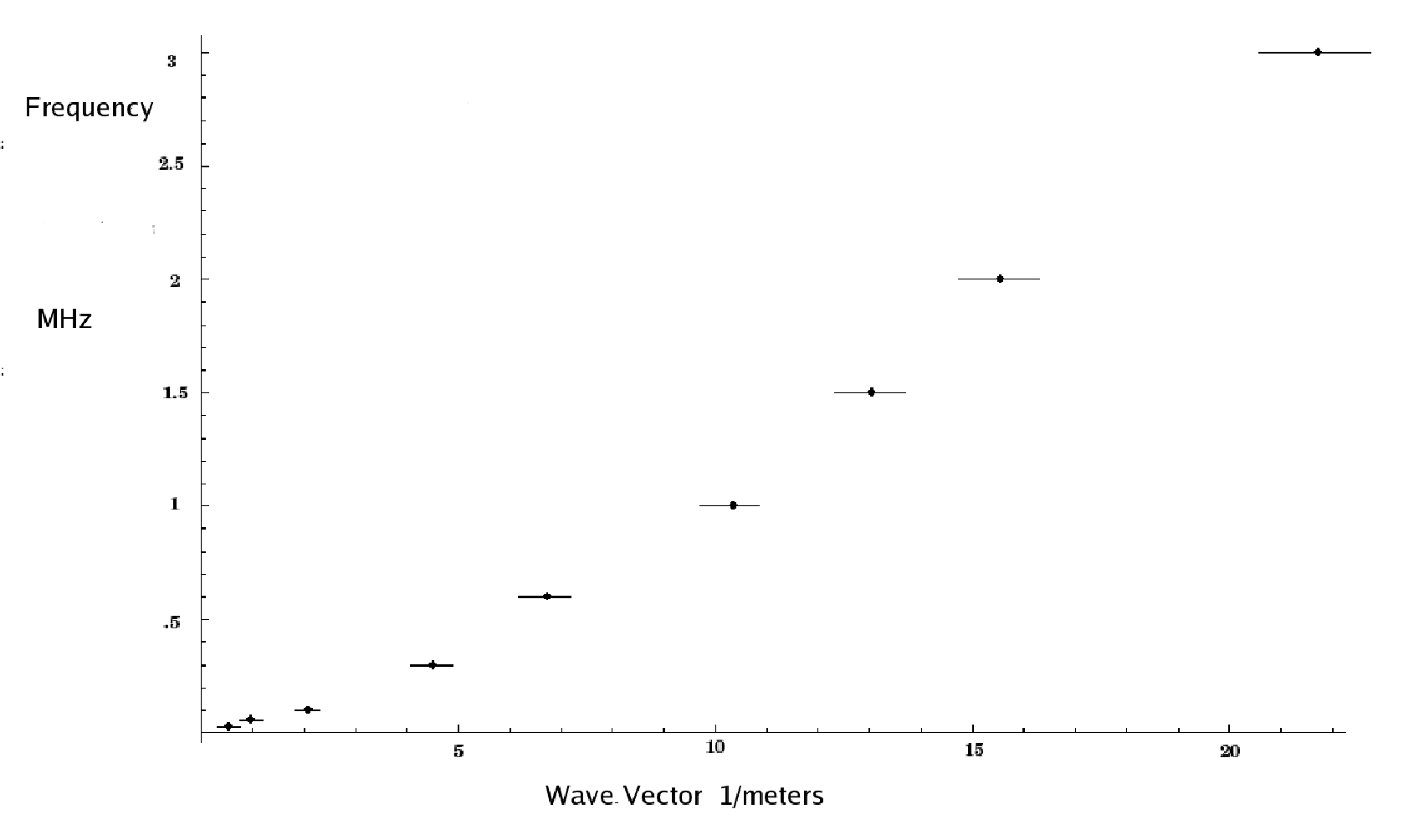}
  \caption{ \textbf{ Field 2 plotted showing a parabolic relation
  between frequency and the propagation vector.  This field was found  to resolvable  in
   the   range from 30 kHz to 3 MHz with both phase and amplitude data.
  }\label{1020 Hot Rolled Steel}}
\end{figure}

\begin{figure}
  \includegraphics[width=6in]{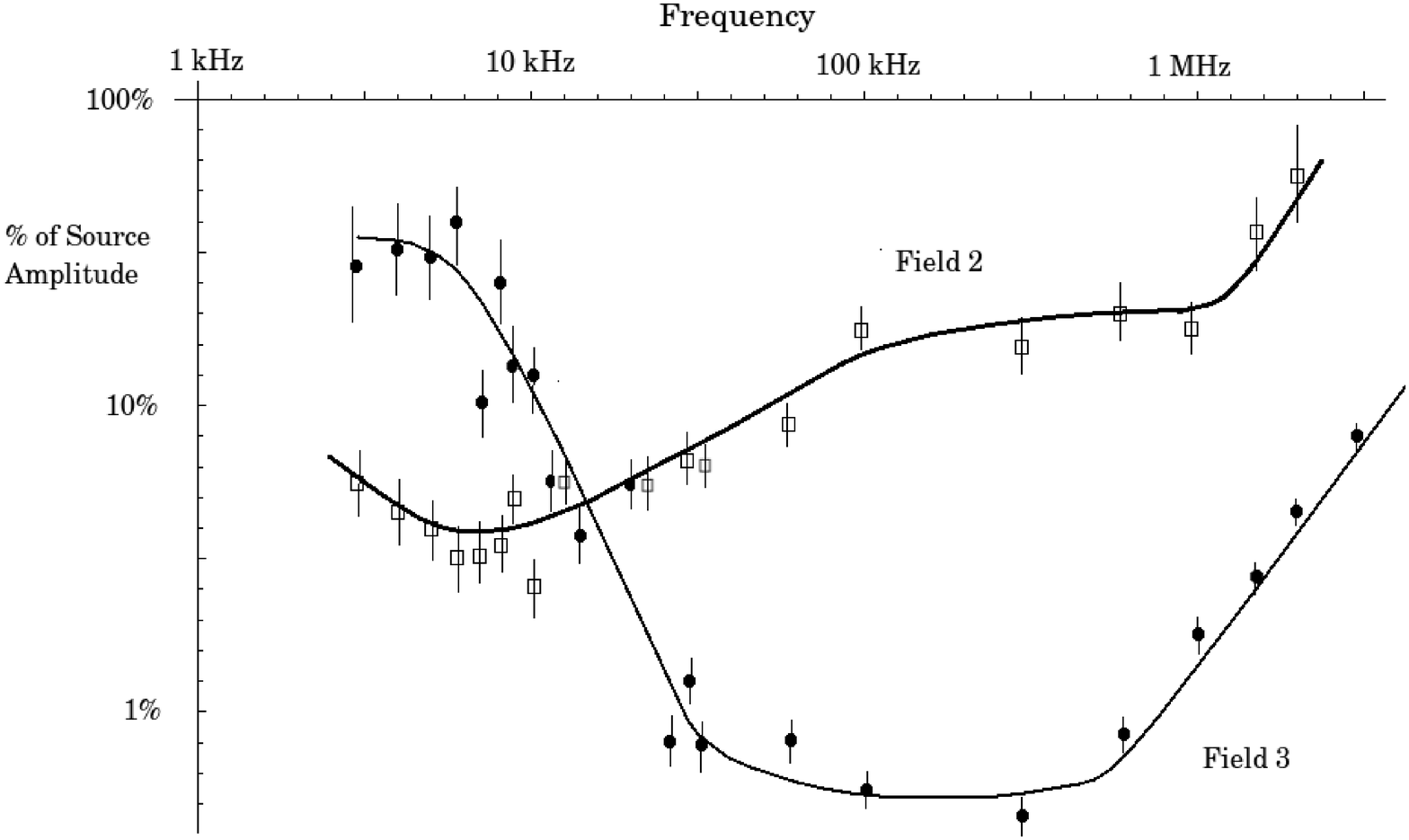}
  \caption{ \textbf{ From the amplitude displacement scans, the fields are
  extrapolated back to zero displacement to
  obtain the $ \% $ of the source  amplitude.  The amplitude of field 1 is not plotted because it cannot
  be unambiguously separated from the source signal. Field 1, even though large at low frequencies, is a
  rapidly decaying short range field.
  }\label{1020 Hot Rolled Steel}}
\end{figure}

At higher frequencies beyond 500 kHz field 3 contribution grows,
figure 14.  This field is at the edge of being resolved at high
frequencies because of its very small phase shift even though it has
a large amplitude.   In the Ln-Ln plot, figure 11,  of the
dispersion curve of field 3 it appears related to field 2 but as a
mirror reflection. This implies that field 3 is really a measure of
the density of a bound state rather than a propagating wave.

  The behavior in the raw dispersion data of field 3 when
amplified above 1.5 MHz is quite complex.   These effects at 3 MHz
are still small but they show both phase and amplitude modulation
patterns that are indicative of new contributions from either the
material or externally detected fields  in combination with the
source fields.

\section{ Mechanisms and Spectra  }

\subsection{ Source Field }

In the transmission displacement scanning, the large signal detected
near the source is assumed to be due to the contribution of
propagating modes that have a radial components.  These components
are local to the source and their influence decreases rapidly as the
sensor moves away from the source. But for those propagating modes
that are rapidly attenuated their overlap with the source fields
inhibits their extraction in this region near the drive.

\subsection{ Field 2}

Of the three sections of dispersion curves found the more familiar
but not necessarily trivial is that of field 2 and will be dealt
with first. The zero field dispersion curve generated from neutron
scattering(20) and all energies are in Joules and wave vectors are
in $ meters^{-1}$  for the thermal spin wave is:

$$   E_{thermal} =  4.5 \times 10^{-40} \textbf{q}^2 J     \eqno[10]  $$

The dispersion curves extracted from figure 13 for the frequency
range of 30 kHz. to 2 MHz. for field   2 is fitted to a simple
quadratic function and yields:

$$   E_{field 2} =   4.69 \times 10^{-30}\textbf{q}^2 J    \eqno[11] $$

The most impressive feature of the function is that its energy is
scaled by a factor of almost $ 10^{10} $ greater than that of the
thermal spin wave dispersion relation.  This field is only detected
unambiguously only in a window from 30 kHz to 2 MHz.  Its amplitude
increase with frequency as its range decreases.  It has an effective
mass of $ 1.8 \times 1)^{-39} kg ( 1.3 \times 10^{-9} m_e ) $. This
is about 10 orders of magnitude greater than that of the thermal
spin wave's effective mass. This very low mass for field 2 makes it
a candidate for a new type of quasi particle or a highly modified
spin wave whose effective mass has been reduced. The question is why
associate this field with that of a spin wave. The first answer is
that it is traveling measurable magnetization with a parabolic
dispersion curve.   The second answer is that the applied induction,
$ \textbf{H}(\omega)$, will lift the phase degeneracy of the group
of spin waves at the applied frequency via the contribution to the
Hamiltonian, $ H_{int} $.

$$  H_{int}   =  - \textbf{m}(t)\bullet \textbf{B}(t)
\eqno[12] $$

This creates a single phase over a short distance that is favorably
to collect a boson concentration.  If this single phase condensation
forms then the exchange scattering interaction of similar frequency
spin waves to exchange phases ceases.  This new allowed state will
then have an effective mass determined by the scattering in its new
state and its collective physical properties will be determine by
its interactions.  These effects in themselves are probably
insufficient to produce such a large change in the wavelength or
mass.

\subsection{ Field 1 }

Field 1 is the first initial sharp decrease in the dispersion
measurements.  This portion of the dispersion measurements represent
a physical region within 2 to 3 cm of the source coil.  This field
is strongly damped and is short range.    Above 30 kHz where the
slope of the decrease was constant as a function of frequency this
probably only reflects the fringing transverse fields from the
source coil.  But from 30 kHz downward  a distinct dispersion curve
can be  extracted. The data for field 1 are not fit by a simple
function that can be extended to zero frequency, if you take the
local slope of the data to extract the velocity of the field values
between 3600 and 5000 $ \frac{meters}{sec} $,  these numbers are not
too different than acoustical velocities.   There is a strong
possibility that field 1 represents a coupled magneto-elastic
wave(21) where a coherent long wavelength spin wave population can
couple to stress wave.  If we compute the phase velocity of field 2
we find that in the frequency range of 3 to 30 kHz.  their values
would run from 400 to 5000 $\frac{meters}{sec}$. This overlap in the
velocities of field 2 with an acoustical wave velocity could allow a
significant coupling to be established. The local dispersion curve
for this coupled stress and spin wave field would be expected to be
strongly varying as the spin wave velocity moved through and past
the acoustical velocity. The wavelength of the field 1 excitation is
between that of the thermal spin wave and the excitation of field 2.
For example at 10 kHz. the wavelength of the field 1 is $ \sim 25 cm
$ whereas the thermal spin wave is 60 microns  and  field 2's
wavelength is on the order of 6 meters. The wavelength of field 1 is
close to the acoustical wavelength at that frequency. The spin waves
that form this coupled magneto-elastic  wave are not from the
thermal spin population but from the long wave length population of
field 2.

When the spin wave velocity of field 2 exceeds that of the speed of
sound the stress and the spin wave  fields should decouple.  The
velocity of field 2 is label $ v_2 $ and is computed below:

$$  v_2  =  \frac{1}{\hbar} \frac{\partial  E_{field 2}}{\partial q}
=  8.89 \cdot 10^4  q  \frac{m}{sec}  \eqno[13] $$

Using the experimentally fitted dispersion curve for field 2 and the
speed of sound in iron at room temperature as 5130 $\frac{m}{sec} $
we find that  $  q > .058 \frac{1}{m} $ this occurs at 30 kHz and it
is precisely where field 2 can start to be unambiguously resolved.

\subsection{Field 3 }

Field number 3 does not look like a traditional dispersion curve as
it has the characteristics of a function describing a simple bound
state. In a coarse examination it is approximately fit by a
quadratic function to the inverse power of 2:

$$   E_{field 3} =  \frac{1.15 \times 10^{-29}}{\textbf{k}^2}  J    \eqno[14] $$

This function   looks qualitatively like a bound state function
where \emph{k} is proportional of \emph{n} the principle quantum
number. That indicates there is probably a significant interaction
energy associated with excitations forming  this state.  This is
particularly emphasized by the increased stability at higher energy
as the size of the state grows larger.  So for field 3 this is not a
dispersion curve but   a measure of the density of the state. The
idea of  bound spin wave states  is not new and was introduced by
Bethe(22) in 1931 for two spin waves on the same band with a net
angular momentum of $ J=2 $.   The coupling there is a simple
parallel spin-spin interaction term.   That interaction does not
look useful in this case. An interaction between two spin waves on
different spin bands  to produce a $ J=0 $ state looks necessary. In
addition, to having zero angular momentum the long wavelength state
 would  weakly interact with the magnetic domain structure. The
reason for considering the zero angular momentum boson is that its
interaction are no longer strongly affected by magnetic domain
orientations.

Placing  the dispersion curves of fields 2 and  density of field 3
in relation with other principle dispersion curves of the EM field
in free space, thermal spin waves and thermal phonon can be seen in
figure 15  which is a schematic representation of the key fields
that will effect detected responses.

\begin{figure}
  \includegraphics[width=6in]{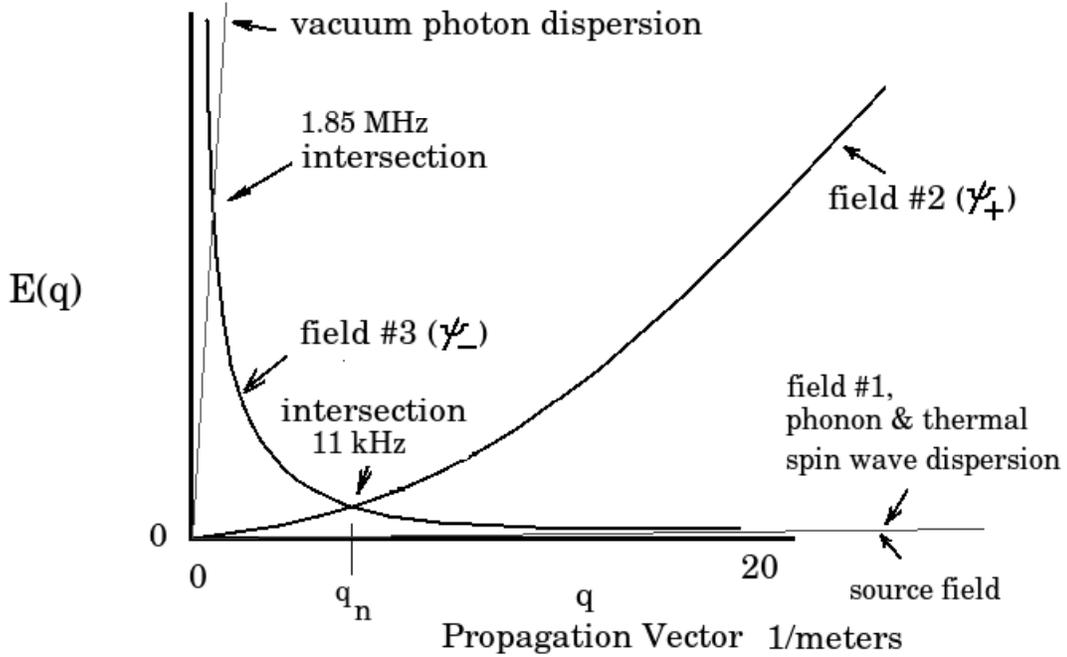}
  \caption{ \textbf{Field 2 and field 3
  are plotted with the vacuum photon dispersion curve.  The intersection points are important as these are regions
  where energy can easily be exchanged between states.   At frequencies greater
  1.85 MHz  the close proximity of the photon dispersion curve
  allows easy coupling between  external fields and  field 3.    Intersection at 11 kHz, labeled $
\textbf{q}_n $ is associated with the
  minimum in transmission
  in figure 12 in the 10 to 30 kHz range. The two fields are degenerate at this crossing. The dispersion curves
  for  field 1, internal transverse EM field, thermal
  phonon and thermal spin waves lie close along
  the $ E \equiv 0 $ axis at this scale of the propagation vector from 0 to 20
  $ \frac{1}{meter}$.  These fields provide a source for conserving
  momentum for transitions
  into or out of states of field 2, 3 and external radiation. }
  \label{1020 Hot Rolled Steel}}
\end{figure}

\subsection{ Populating a BELC with a Single Phase }

High temperature Bose-Einstein like condensations (BELC) have been
detected in spin wave systems that are laser pumped(23).  The reason
for differentiating this condensation from that of a BEC is that
these are pumped at a frequency greater than zero.   The expression
for the transition temperature(25) to the BEC is :

$$  T_{BEC}  =  \frac{\textbf{n}^{2/3}}{m}\frac{2\pi\hbar^2}{k_b \zeta^{\frac{2}{3}}(\frac{3}{2})}  \eqno[15] $$

Where \textbf{n} is the number of spin waves,  m is the mass, $\zeta
$ is the zeta function.   To have a $ T_{BEC} = 770^o$C  one only
need an occupation number of   greater than \textbf{n} $=
5.78\times10^{12} $ and at $23^o$C the required density is $ 8.73
\times 10^{11}$ .

To compare with the thermal spin wave density at 10 kHz  one has
density $ ~ 6 \times10^8 $ at $ 23^o $C and $ 2 \times 10^9 $ at $
770^o$ C. This indicates that forming BEC at either of these
temperatures is only favorable if you pump the state with a non
thermal source to form the state.

\section{ Wave Equation for the BELCs }

The basic form of wave function for a BEC is the Gross-Pitaevskii
equation(25)   for the ensemble as a whole for a constant number of
bosons is:

$$ i\hbar\frac{\partial \psi(\vec{r})}{\partial t}  = (
-\frac{\hbar^2\nabla^2}{2m} + V(\vec{r}) +U_o|\psi(\vec{r})|^2)
\psi(\vec{r}) \eqno[16] $$

The induction problem considered here is a steady state problem
where the number of spin waves is fixed by the drive amplitude. The
sign of the potential $ U_o $ along with scattering determines the
lifetime of the condensate. The solution of the equation with no
external potential in the limit of low density reduces to the
Schr$\ddot{o}$dinger equation.  Simplifying  further assuming no
binding potential to the matrix and  the cyclical boundary value
conditions are dropped  the solutions  then are not restricted to
Block functions. The solutions can have wavelength greater than the
sample size. By examining deviations in the band structure from free
particle motion, $U_o(q)$ sign and magnitude could be extracted
which contain the interaction of the elements in the BELC.  In the
limit of a low density population of BELC spin waves ~ $ 10^{12} $
the term can be dropped for trial solution.  The external potential,
V(r),  has a value in the region of the source coil. The internal
ferromagnetic magnetization $\textbf{M}(x,y,z)$ is a rapidly varying
function because of the domain structure and its integral over the
volume of the sample will vanish as the size of the BELC structures
are large relative to the MDBs. So that the Gross-Pitaevskii
equation is just reduced to a free particle Schr$\ddot{o}$dinger
equation away from the the sources and picking a linear one
dimensional geometry of figure 1, considering the solutions along
the z-axis where the length of the bar is from $ \frac{-L}{2}$ to $
\frac{L}{2}$ .

$$ i \hbar \frac{\partial \psi(z)}{\partial t}  = -\frac{\hbar^2\nabla^2}{2m}\psi(z) \eqno[17] $$

This simple equation has   solutions made up from the products $
e^{\pm iqz} $,  $ e^{ \pm a z } $  multiplied by $ e^{\pm i \omega
t} $ where  $ a $ is a real number.   Taking the two positive energy
solutions, the exponentially decaying solution as $ \psi_- $ and the
plane wave as $ \psi_+ $.    Our choice is helped from the
experimental dispersion curves of figure 11.   The parabolic curve
looks very much like a simple free particle:

$$ \psi_+(z,t)  = \frac{1}{\sqrt{L}} e^{\pm iqz - i\omega t}  \eqno[18] $$

The hyperbolic dispersion curve for field 3 in figure 11  is even
more interesting   $ E(k) \sim \frac{1}{k^2} $. This allows us to
set  $a  = \frac{\alpha}{k} $ and then solution is a decaying
exponential, where $ \alpha $ has units of $ \frac{1}{meters^2}$.
This function describes a linear magnetic polarization of the medium
coupled to the source which is not propagating. These are steady
state responses and  looks like  an oscillating linearly polarized S
state bound to the source.

$$ \psi_-(z,t)  =  \sqrt{\frac{ \alpha }{k( 1 - e^{\frac{-L\alpha}{k}})}} e^{-\frac{\alpha z}{k}  + i \omega t}, z > 0     $$
$$ \psi_-(z,t)  =  \sqrt{\frac{ \alpha }{k( 1 - e^{\frac{-L\alpha}{k}})}} e^{\frac{\alpha z}{k}  + i \omega t}, z <0  \eqno[19]  $$

The $  \psi_-(z,t) $  state has a linear momentum  $ <p> = 0 $
unlike the propagating mode.  This zero momentum solution is
analogous to the zero momentum state of gas phase BEC.    As the
frequency is increased the size of the state is increases. This is a
feature that dominates at higher frequencies. The two dispersion
curves are

$$ E_+(k)  =   \frac{\hbar^2q^2}{2m}   \eqno[20] $$

$$ E_-(k)  =   \frac{\hbar^2\alpha^2}{2mk^2}  \eqno[21] $$

 These two solutions  are driven by the applied field
 are easily resolved in the frequency band 30 kHz. to 3mhz. The
second solution has some interesting characteristic at high
frequency the field decays very slowly spanning large distances
relative to the source. The intersection point between the two
dispersion curves only depends on the corrections to the effective
mass as a function of their propagation vectors.  The parameter $
\alpha $ can be found from the crossing of the curves for field 2
and 3 from figure 11 at $ q_n $  then $ \alpha = q_n^2 $ assuming
the effective mass of each state is the same.

The wave function has to be defined for the spin band on which the
spin wave is moving.  So that in the case of iron where there are
two available bands the wave function would be in a vector form:

$$ \psi_\pm   =   \begin{pmatrix}  \psi_\pm^\uparrow \\ \psi_\pm^\downarrow
\end{pmatrix} ,  iron    \eqno[22]  $$

Originally the spin wave was proposed as an excitation  to reduced
the magnetization of the aligned ground state of a ferromagnetic
material.  For the right handed or left handed polarized spin wave
on a single band the angular momentum was J=1 and it is opposed to
the magnetization vector of the band.  For a material with a net
magnetization there is only one circular polarization active.  This
excitation has three components of time dependent magnetization $
\textbf{m} = ( \theta,
 \theta + \frac{\pi}{2}, \theta + \pi) $ equivalent to the magnetic quantum number
 m.  Since within each magnetic domain there is a strong net moment
 only one circular polarization will be active on each band. This has its origins
 in the torque experienced by the band moments via $ \dot{\textbf{M}} = g \textbf{M}\times \textbf{H} $.
  For
 the spin up band  this will be the right hand circular polarization
 ( \textbf{RH}$\uparrow$ ) and for the spin down band this the left hand
 circular polarization ( \textbf{LH}$\downarrow$ ).
Within any magnetic domain these spin waves exist and interact
strongly with   magnetic domain boundaries (26). On a single band
spin wave coupling of two spin waves will produce a higher angular
momentum state, $J=2$. With two available bands of opposite spin
orientation, the coupling of two spin waves one on each band can
produce a, $J=0$, state with an axial time dependent magnetization.
Considering the propagation vectors of the individual spin wave
\textbf{q}, the allowed orientation of two spin waves on opposite
bands fall into the range $ 0 \leq |\textbf{q}_{sum}| \leq 2
|\textbf{q}| $.   In this range there are only two state that will
be persistent.  These are $ \textbf{q}_{sum} = 0 $ and the $
\textbf{q}_{sum} = 2 \textbf{q} $. For other value of
$\textbf{q}_{sum}$  the pair becomes non local and unbound.   The
 $J=0$ state is a very different in that across magnetic domain
boundaries the strong force due a finite angular momentum with the
changed spin alignments  no longer operate.   Also for this spin
zero boson, the scattering between degenerate phase states is also
removed as they don't exist for the spin zero pair.    With the
magnetic domain boundaries becoming transparent to the $J=0$ states
the change in physical properties of these coupled spin waves will
be altered by operating across the entire sample.

\bigskip

\textbf{Table 5: Pairing of spin waves on two bands where the total
angular momentum is $J=0$. The right hand(\textbf{RH}$\uparrow$) and
left hand(\textbf{LH}$\downarrow$) polarizations are referenced to
the orientation of the net local moment within a magnetic domain.
The arrows refer to the specific band on which the spin wave is
active. Because of a strong net moment only the
\textbf{RH}$\uparrow$ and the \textbf{LH}$\downarrow$ polarizations
are active on their specific bands.  The value $ \theta $ indicates
 the spin wave phase relative to the local applied time
dependent field, and the orientation is the angle between the linear
polarization of the magnetization and the axis of the time dependent
field.}

\begin{center}
  \begin{tabular}{|c|c|c|c| }
  \hline
\textbf{Band}  $ \uparrow  $ and \textbf{Phase} & \textbf{RH}$\uparrow$  $\theta = 0^o $ &  \textbf{RH}$\uparrow$ $\theta = 0^o $  &\textbf{RH}$\uparrow$ $\theta = 0^o $ \\
\textbf{Band}  $ \downarrow  $ and \textbf{Phase} & \textbf{LH}$\downarrow$ $\theta = 0^o $ &  \textbf{LH}$\downarrow$ $\theta = \pi $  &\textbf{LH}$\downarrow$ $\theta = \frac{\pi}{2}  $\\
\hline

\textbf{Polarization}  &  \textbf{Linear} & Linear & Linear  \\
\textbf{Orientation}   &   $\textbf{0}^o $ &  $\frac{\pi}{2} $ &   $\frac{\pi}{4} $ \\

\hline
  \end{tabular}
\end{center}
\bigskip

The only phase that is energetically favorable with respect to a
time dependent source field is the first column where both the left
and right hand components are in phase with the applied field. The
second column is also worth noting that if the phase of one
component is shifted by $ \pi $ during an interaction then the
orientation of the magnetization rotates by $\frac{\pi}{2}$ which is
an easy condition to detect.

There are really two regions where the time dependent vector
potentials needs to be consider. First is the region under the
source coil and second in the remainder of the bar. These fields are
very different. The induced source  field is a rapidly decaying
transverse field limited to the material near surface  directly
under the inductor. The field due to the long wave length spin waves
is a very slowly spatially varying field. The wave length of the
magnetization signals   are long relative to thermal spin waves, the
source appears almost as a delta function source at the location of
the coil.  This source function is given by the interaction
Hamiltonian, $ H_{int} $, which is the product of the applied field
with the time dependent magnetization. Away from the source the time
dependent magnetization of the BELC provides a field energy that is
dependent on the interaction between the created BELC components.
Because the curl of this BELC driven field is small, eddy current
loses will be small. An example of these fields slowly varying
distributions probed with a small sensor over a larger bar than can
be found in reference (27) where a field produced from BELC
interactions generate a uniform field  across the bar away from the
source.

The  two fields isolated above 30 kHz represent a propagating mode,
$ \psi_+ $ which has a real momentum and   a bound state, $ \psi_- $
with zero net momentum. The transmission experiment where the center
of the bar was raised above the Curie point,  shows a drop in
transmission with an abrupt decrease in phase delay. From the
dispersion relations at frequencies greater than 1 kHz the phase
delay at a fix offset is less for $ \psi_- $ than for $\psi_+ $.
Considering this phase property, $\psi_+ $ looks filtered or
reflected out and $ \psi_- $ has managed to penetrate this
paramagnetic region without significant attenuation.

\subsection{ Coupling Spin Waves to Form $ J=0 $ Pair}

The wave equation solution for the spin wave excitation, $\psi_+ $
is a good representation for the measured field 2 of figure 11.
However field 3 is represented by, $ \psi_- $ is a decaying spin
wave state that dominates at high frequencies and appears to be able
to traverse  regions above the Curie point.  The question is what
can drive the formation of a $ J=0 $ state which is quite unlike the
$J=2$ coupled state for spin waves on the same band.  These long
wave length excitations are measured a sample of finite size much
smaller than their wavelength. In an isotropic material one can
calculate the magnetic field energy of two separate configurations
bound or separated spin waves where the symmetries and phases are
matched to produce and axial magnetization of the paired state as
defined in the previous section.

  $$  H_{separate}  =  ( 0, cos(\omega t)H_1, sin(\omega t)H_1)
   \times  2
 \eqno[23] $$

  $$  H_{pair}    =    ( 0, 0 , 2 sin(\omega t ) H_1 )  \eqno[24] $$

computing the squares of the field and multiplying by the vacuum
permeability, $\mu_o $.

$$  E_{separate}  =  -   \mu_o H^2_1  \eqno[25] $$

$$  E_{pair}   =   - \frac{\mu_o}{2}  < 4sin^2( \omega t) H^2_1 > = -  \mu_o H^2_1  \eqno[26] $$

The field energies for the two configuration are identical and there
is no driving force for forming a bound pair.  The calculation for
strain energy due to the spin wave field is similar showing equal
contribution  to both configurations.   However, iron bases alloys
are   magnetically anisotropic  with $ \sim 30\% $ increase in the
permeability(28) along the easy magnetization cube axis, $ <100> $.
The orientation dependent strain energy contribution is going to be
neglected.  The bulk of the spin waves energy lies in its magnetic
field this anisotropy gives as reduction in energy for the linearly
polarized magnetization on the easy axis of $ \sim
-\frac{\hbar\omega}{3} $, which is a large interaction energy
favoring the formation of coupled spin wave pairs.  So that texture
and grain orientation can play a large roll in the ability to form
the paired state BELC. To calculate the effect of the anisotropy on
the binding energy, the random orientation of a collection of grains
with projections on the easy axis and the hard axis have to be made
on a line.   For any line segment taking a random selection of the
three orientations,  $ \frac{1}{3} $ falls on an easy axis. This
occurs for both the paired and separate states so there is no
driving for binding a paired state.   However, if the hot rolled
textures the bar tested was preferentially $ <100> $ along the long
axis then the paired states would be preferential.   Orientation
texture comes from two source, the original as cast grain
orientation columnar growth into the melt is typically $ <100> $. In
continuous cast product this is not the case for the entire billet
cross section. The casting growth proceeds from the wall of the
casting and from the bottom in the middle portion of the casting so
that the cast texture is mixed. The second source is from secondary
recrystallization after cold rolling and this is tied to minimizing
surface energy of a sheet which is not the case for this bar.
Oriented single crystal of iron, and cold rolled sheet  should show
these effects but not hot rolled rod.

The mechanism  that is  active in driving the binding of spin pairs
in poly crystalline material arises from the minimization of leakage
field at the surface.  The energy of the bound state will be
decreased because the fields normal to surface of the material will
vanish.  The leakage fields contribute to this higher energy for the
unpaired spins can be seen by replacing $ \mu_o $ with $ \mu $ for
the \textbf{z} component of the field in the limit of a very thin
rod when $ \mu > \mu_o $.

$$  E_{separate}  =  - \frac{\mu + \mu_o}{2}    H^2_1  \eqno[27] $$

$$  E_{pair}   =   - \frac{\mu}{2}  < 4sin^2( \omega t) H^2_1 > = -  \mu H^2_1  \eqno[28] $$

$$  E_{pair}   <   E_{separate}     \eqno[29] $$

This rough calculation is valid when $ \mu > \mu_o $ for the
energetics of the $ J=0 $ paired states as compared to two
separately located spin waves making pairing favorable at all
frequencies when driven with a time dependent field.   This would be
true for either the non propagating $ \psi_- $ or the propagating $
\psi_+ $ field.   This argument can be generalize for an
inhomogeneous material which contains magnetic domain boundaries,
secondary phases such as oxides, martensite, austenite, carbides or
hydrides.  Writing the permeability as:

$$    \mu_{ij} =  \frac{\partial B_i}{\partial
H_j} = \sum_k \frac{\partial B_i}{\partial x_k}\frac{\partial
x_k}{\partial H_j}  \eqno[30]  $$

Considering those regions where \textbf{H} is slowly varying  and $
\frac{  \partial x_k}{\partial H_j }  \neq 0  $ .   The quantity $
\frac{\partial B_i}{\partial x_k}  $  carries the measure of the
inhomogeneity in the permeability.   It is this quantity that
reflects $\mu_{ij}(\textbf{x},\omega)$ variation in the material. If
$\mu_{ij}(\textbf{x},\omega)$ is not a constant function over the
material then there will be an energetically favorably coupling
between spin waves to form axial magnetization pair with  $J=0$ if
the band structure allows such a state.  The strength of the
coupling will depend on the integral of $ \frac{\partial
B_i}{\partial x_k}  $ over the volume which can take on large values
traversing the magnetic domain boundaries and second phase
precipitates.

Reversing the picture again and having ferromagnetic precipitates in
a non ferromagnetic matrix is much like the first simple model
presented above.    Some non magnetic matrix that have embedded
ferromagnetic material seem to be able to support the linear time
dependent magnetization of these bound spin wave fields with minimal
loss. This becomes important in the non destructive testing of
stainless steel welds by inductive techniques where the heat
affected zones and the weld zones contain a small minority of
ferromagnetic precipitates whose responses can often mask large
physical defects.

\subsection{ \textbf{ Coupled Spin Waves Interactions and BELC Stability }}

\bigskip
One feature of a stable BEC is that the particles involved should
show a repulsive interaction if the state is to be easily formed.
This interaction can be computed from the density variation as one
considers the displace of two particles by an amount $ \Lambda $.
There are three different cases of interest.  The first two are
inter particle interaction between states of the same type.  The
third case is the interaction between a displaced  $ \psi_{+} $ and
a $ \psi_{-} $ state.  In the case of large sample size relative to
the wave length the plane wave solution $ \psi_{+} $  produces no
net interaction. However, in truncated sample the net interactions
are altered.   If the wave length is long then one can can take a
trial function $ \psi_{+}  = e^{ i( \textbf{q} \textbf{x} - \omega t
)} $. The wave functions of equation 18 and 19 can be expanded for
the near field behavior as the bar length, $ L $, is a highly
truncated space:

$$   \psi_{+} \approx  \frac{1}{\sqrt{L}} e^{  -i\omega t} ( 1 + i\textbf{q}\textbf{x} ) \eqno[31] $$

Similarly this can be done for the $ \psi_{-t} $  in the case of
state that is larger compared to the bar length:

$$   \psi_{-} \approx \frac{1}{\sqrt{L}} e^{  +i\omega t} ( 1 -  \frac{\alpha x}{k} ) \eqno[32] $$

Now the time dependent magnetization must be added to the system for
the two bands and it has to absorb the time dependency  explicitly.

$$  \psi_{\pm  } \approx     f(\textbf{x},\textbf{q})
( \textbf{M}^\uparrow(t)  + \textbf{M}\downarrow(t) ) \eqno[33] $$

Where for a spin one boson the magnetization for  one band are
represented by three states relative to the phase of the local
source field $ \theta $ and the bold face state is in phase with the
source field within the sample and in the coupled coherent state on
the in phase spin wave field component remains:

$$  \textbf{M}(t,\theta)   = \begin{pmatrix}  \textbf{m}(t,\theta)  \\  m (t,\theta + \frac{\pi}{2} )\\
 m (t,\theta +  \pi )
\end{pmatrix}   =  \begin{pmatrix}  \textbf{m}(t,\theta)  \\ 0\\
 0
\end{pmatrix}  \eqno[34] $$

$$   \begin{pmatrix}  \textbf{m}\uparrow(t,\theta)  \\ 0\\
 0
\end{pmatrix}  +   \begin{pmatrix}  \textbf{m}\downarrow(t,\theta)  \\ 0\\
 0
\end{pmatrix} \Rightarrow (  \textbf{m}\uparrow(t,\theta) +
\textbf{m}\downarrow(t,\theta) ) \eqno[35] $$

The $J=0$ paired  function  $  \psi_{-} $ and $ \psi_{-} $  become:

$$  \psi_{+} \approx \frac{1}{\sqrt{L}}   ( 1 + i\textbf{q}\textbf{x} )
  ( \textbf{m}^\uparrow(t,\theta)  + \\
 \textbf{m}^\downarrow(t,\theta) )
     \eqno[36] $$

$$  \psi_{-} \approx \frac{1}{\sqrt{L}}    ( 1 -  \frac{\alpha x}{k}  )
  ( \textbf{m}^\uparrow(t,\theta)  + \\
 \textbf{m}^\downarrow(t,\theta) )
     \eqno[37] $$

It is possible that the large effective mass of the thermal spin
wave has its origin in the interaction of the six $
\textbf{m}\uparrow\downarrow(\theta)$ states with each other that
are populated in addition to the magnetic domain boundaries.  If
there were higher angular momentum states with coupling on the same
band at the same wave vector the number of interacting states would
increase.   The $J=0$ states allow the functions to span the
magnetic domain boundaries and removes the scattering into other
degenerate phase states which no longer exist.

\subsection{ \textbf{Two Particle Interactions  } }

Now there is simple calculation that can be done on both wave
functions to determine   the sign of the inter particle potential
interaction if the one wave functions is  displaced a distance , $
\Lambda $. The bar is a truncated space on which a set of slowly
varying wave function are considered. Using the data in figure 1
that the potential will be a minimum for the higher density of
coherently coupled spin waves. This makes the sign negative on the
integrals for a density increase over the bar to scale   an
interaction energy.

$$ \delta\rho_{++}(\Lambda) \sim \frac{1}{L} \int_{-\frac{L}{2}}^{\frac{L}{2}}
( \psi_{+}^*(0)   +  \psi_{+}^*(\Lambda)  )( \psi_{+}(0) +
\psi_{+}(\Lambda) )dx \eqno[38]    $$

$$  \psi_{+}(\Lambda)  =  \frac{1}{\sqrt{L}} ( 1 + iq( x - \Lambda )
\eqno[39] $$

Just taking the term with a factor of the separation $ \Lambda $
where:

$$   \delta\rho_{++}(\Lambda) \sim  q^2 \Lambda^2   \eqno[40] $$

The potential as a function of $ \Lambda $ is proportional to the
negative of this density :

$$ V_{++}(\Lambda) \sim   - q^2  \Lambda^2    \eqno[41] $$

The inter particle interaction is repulsive and increasing in
strength  with frequency for this excitation.

For the case of the  $ \psi_{-} $  excitation   the resulting
integral has to be done over three ranges of over lapping functions:

$$ \delta\rho_{--}(\Lambda) \sim \frac{1}{L} \int_{-\frac{L}{2}}^{\frac{L}{2}}
( \psi_{-}^*(0)   +  \psi_{-}^*(\Lambda)  )( \psi_{-}(0) +
\psi_{-}(\Lambda) )dx \eqno[42]    $$

Where

$$  \psi_{-}(\Lambda)  =  \frac{1}{\sqrt{L}} ( 1 - a( x - \Lambda )
,  x > \Lambda  $$

$$  \psi_{-}(\Lambda)  =  \frac{1}{\sqrt{L}} ( 1 + a( x - \Lambda )
,  x < \Lambda \eqno[43] $$

computed over the three regions of overlap  the result is:

$$ \delta\rho_{--}(\Lambda)  \sim  a^2\Lambda^2  + 4a\Lambda\frac{\Lambda}{L}( 1 - \frac{a\Lambda}{3})   \eqno[44] $$

This  results in a potential dependent on the separation $ \Lambda $
which is interesting in that it can be both attractive and repulsive
depending on the length of the sample and the frequency.

$$  V_{--}( \Lambda)  \sim   - a^2\Lambda^2  - 4a\Lambda\frac{\Lambda}{L}( 1 - \frac{a\Lambda}{3})\eqno[45] $$

The interaction cannot be attractive unless:

$$  \frac{a\Lambda}{3} > 1 \eqno[46] $$

which can only occur at high frequencies.

The third interaction is that between  $ \psi_{-} $ and   $ \psi_{+}
$ and that can be computed with the integral:

$$ \delta\rho_{+-} (\Lambda) \sim \frac{1}{L} \int_{-\frac{L}{2}}^{\frac{L}{2}}
( \psi_{+}^*(0)   +  \psi_{-}^*(\Lambda)  )( \psi_{+}(0) +
\psi_{-}(\Lambda) )dx \eqno[47]    $$

The resulting density dependence on the separation is:

$$ \delta\rho_{+-} (\Lambda) \sim  a^2\Lambda^2( 1  - \frac{4}{aL})
\eqno[48] $$

This interaction can only go attractive at low frequency when $
\frac{4}{aL} > 1 $   or if the sample size is small.

\section{ Spectroscopy of the BELC and  Induction Analysis }

The kinetics of pumping  BELCs are controlled by the density of
mobile magnetic domain boundaries.  At the magnetic domain
boundaries the magnetization transitions through zero allowing the
applied source fields to drive transitions.  The principal
transition will be to drive the ground state, aligned ferromagnetic
ordering, into a time dependent state. The applied field will couple
the vacuum state into $\psi_+ $ or $ \psi_- $.

The fundamental difference between a BELC   and a BEC  is the time
dependence associated with the BELC  and it has to be continuously
pumped as a steady state process at a finite frequency $ \omega > 0
$. The easiest way to explore this is with a perturbation. These
states are not easily perturbed, there are few structures that alter
their characteristics.  The simplest measurable perturbation is to
induce a second perturbing BELC where we can alter its level. Going
back to figure 3b of our transmission experiment and    changing the
geometry slightly by replacing the heat injected into the center of
the bar with a low frequency field time dependent field $
\textbf{H}(\omega_2 t)$. A coil of the same type will be used to
inject the center field.  By displacing the two fields by a 5
centimeters we are insuring that we do not have a direct
superposition of our two applied fields.

Our original field injected at the end will be $\omega_1 $ driven at
the same levels $ \textbf{H}(\omega_1 t)$ as our transmission
experiments. The fields are driven under these conditions, $
\omega_2 \ll \omega_1 $ and $ |H(\omega_2)|
> |H(\omega_1)| $.  The spacing separation between the source and
the receiver is set at 10 cm.  As the field level $ H(\omega_2) $ is
increased two states emerge,  the strongest being  $ | \omega_1
-2\omega_2 > $ followed by $ | \omega_1 +2\omega_2 > $ and if the
field level is increased a little more another pair of states
develop with the strongest being $ | \omega_1 + \omega_2 > $
followed by $ | \omega_1 - \omega_2 > $. At this stage the original
probe field $ | \omega_1>$ has managed a gain in amplitude of about
4$\%$.  A spectra of these are shown in figure 16.

\begin{figure}
  \includegraphics[width=6in]{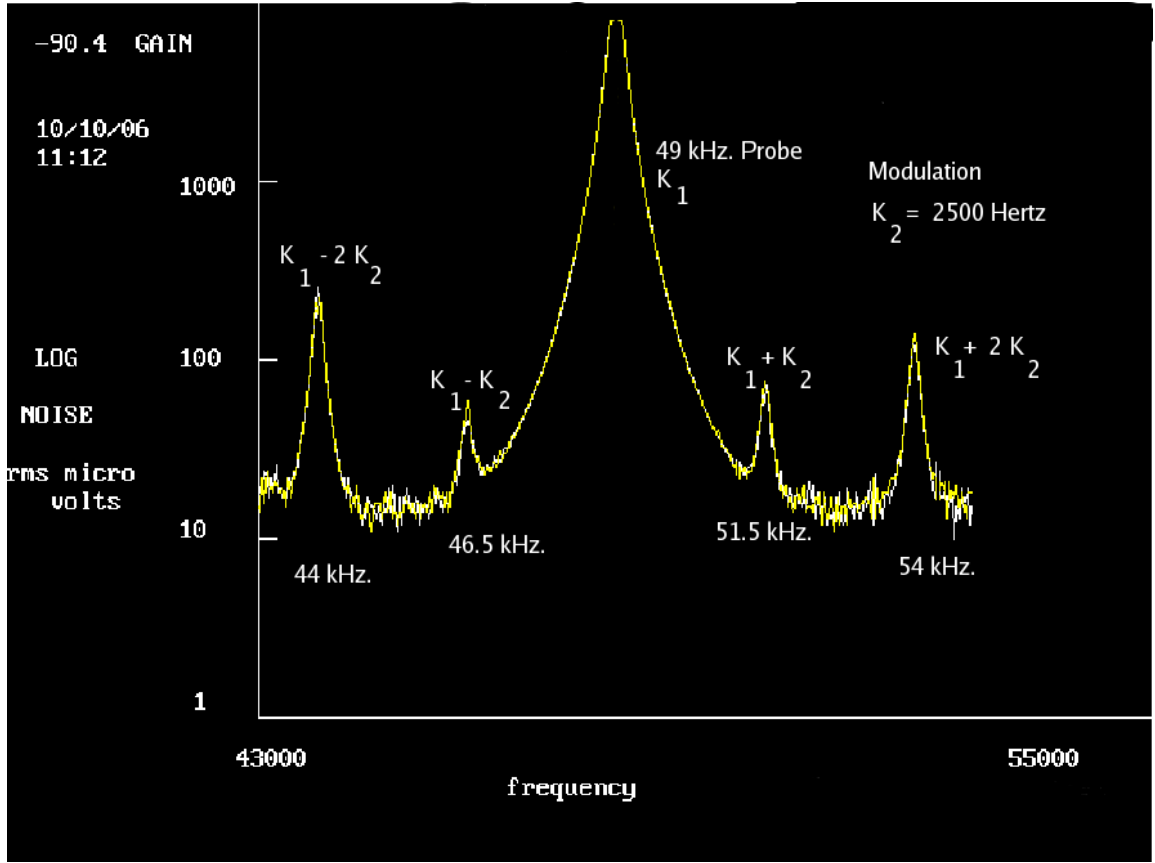}
  \caption{\textbf{The 4 principal transitions observed when two BELCs overlap. $K_1 $ is the
   wave vector for the weaker probe field $ \omega_1 $  and $K_2$ is
   the wave vector for the stronger drive field $ \omega_2 $  injected at the
   center of the bar.   All signals are detected at the opposite end of the bar that is 10 cm away. }}
  \end{figure}

If you keep increasing the drive level of $ \omega_2 $ you generate
a spectral comb of states $ | \omega_1 \pm n \omega_2 > $ where n is
an integer. The even number states and odd number states differ in
how the transitions alter the polarization of the magnetization
relative to the propagation vector so that apparent strengths of
transitions are dependent on the geometry and the interaction that
are active when measured. There are other measurement geometries
more suited to working out the details of the individual
transitions. The full set of coupling from table 6 need to be
considered  as possible contributors to this spectra.

There is a possible fault in this view of the spectra. In figure 4
the single coil response shows a marked nonlinear behavior for iron
and because of this one would expect to generate the product and
difference fields when two fields are superimposed on the material.
In this case the introduced fields are physically displaced as is
the detector. Second the spectra generated are more characteristic
of  linear and angular momentum  conserving reactions forming their
own BELC as they are then found at a remote  detector. Simply having
a local nonlinear field response does not account for their
migration to the detector or the details of the spectra.

\bigskip

\textbf{Table 6:  Relative transition strengths with a probe field $
\omega_1 $  of 49 kHz.   The strengths are taken at the extrapolated
limit for the injected field $ \omega_2 \rightarrow  0$.  }
\begin{center}
  \begin{tabular}{|c|c|c| }
  \hline
\textbf{Transition } &\textbf{ Strength = $ \% $ of $ <\omega_1|\omega_1> $ } \\
\hline
$ <\omega_1| V_1 | \omega_2 > $ &                    4   $\pm$ .1        \\
$ <\omega_1 - 2 \omega_2 | V_2 | \omega_1 > $ &       2.4   $\pm$ .1         \\
$ <\omega_1 + 2 \omega_2 | V_2 | \omega_1 > $ &       1.4   $\pm$ .1        \\
$ <\omega_1 +   \omega_2 | V_3 | \omega_1 > $ &        .9   $\pm$ .1          \\
$ <\omega_1 -   \omega_2 | V_3 | \omega_1 > $ &        .8   $\pm$ .1       \\
\hline
  \end{tabular}
\end{center}
\bigskip

\subsection{ Hamiltonian for the Transitions }

The kinetic portion of the Hamiltonian for the propagating spin wave
excitation neglecting their mutual and external field interactions
can be written as a free particle is :

$$   H =\frac{1}{2m_{s-w}}\sum_j
\hbar^2 \textbf{q}^2_j  \eqno[49]
$$

In a material like iron where the conduction electrons are also
participating in the spin wave excitation the Hamiltonian will
include the kinetic terms for the electron.  The electrons  in turn
are  acted upon by any field present. The effect of the slowly
varying vector potential which is only due to the BELC's themselves
should be a break on the momentum(29) where $  \textbf{p}
\rightarrow \textbf{p} - \frac{e}{c}\textbf{A} $.   Since the
conduction electrons supporting the spin waves are acted on by the
same field the momentum of the spin waves will also be altered by
the external vector potential. Therefore the same transformation, $
\hbar\textbf{q} \rightarrow  \hbar\textbf{q} - \frac{e}{c}\textbf{A}
$ , for the field dependent momentum has to be made for a material
like iron. The total wave function which will include the spin wave
contribution will act on the two field dependent terms of the vector
potential. The velocity of the fields within the conductor will not
be speed of light in vacuum but will be the phase velocity of the
particular component of the vector potential that is being
considered.   The induced transverse field will have a substantially
reduced phase velocity equal to $ \sqrt{\frac{\omega}{\mu\sigma}} $.
This will also amplify transitions driven by these fields.

$$   H =
\frac{1}{2m_{s-w}}\sum_j ( \hbar \textbf{q}_j
-\frac{e}{c}\textbf{A})^2 \eqno[50]
$$

With the mass of the electron so much greater than that of the BELC
spin wave the two field dependent terms in the spin wave portion of
the kinetic energy get amplified in their ability to drive
transitions by a factor of $ 10^9 $.

 The vector potential
$\textbf{A}$ is actually considered in two parts, that of the source
field, $ \textbf{A}_{ext} $   and the vector potential due to a
specific BELC field  $ \textbf{A}_i $.   The external field,  $
\textbf{A}_{ext} $, is a locally acting field and
 $ \textbf{A}_i $ acts over the entire volume.  They arise from

$$  \textbf{B} =  \mu_o \textbf{H}_{ext} + \sum_i^{BELCs} \textbf{M}_i =   \nabla \times \textbf{A} \eqno[51] $$

where

$$  \nabla \times \textbf{A}_i  =  \textbf{M}_i  \eqno[52] $$

$$    \textbf{A}   = \textbf{A}_{ext}  +  \sum_i^{BELCs}
\textbf{A}_i  \eqno[53] $$

These two fields have very different spatial distributions.    In
the computation of the induced vector potential field $
\textbf{A}_{ext} $ at the source within a homogenous conductor the
solution is a single, inward propagating, exponentially decaying
field.  This correct solution actually disguises what is taking
place. There are in fact two field $ \overrightarrow{\textbf{A}} $
and $ \overleftarrow{\textbf{A}}  $ along with circumferential
current $ \textbf{j}_\theta $.  One field propagating inward  $
\overrightarrow{\textbf{A}}  $  and a reflected field  $
\overleftarrow{\textbf{A}}  $.   The quadratic then becomes:

$$   \textbf{A}_{ext}^2 =  \overrightarrow{\textbf{A}}^2 +
\overleftarrow{\textbf{A}}^2 + 2
\overrightarrow{\textbf{A}}\overleftarrow{\textbf{A}} \eqno[54] $$

The third term is the interesting term in that it carries no linear
momentum.    A two photon absorption into the vacuum state can
produce a pair of spin waves with net zero linear momentum while
conserving angular momentum and energy for a two spin wave
transition. This break down of the vector potential only occurs in
  conductors. The reflection response in nonconducting ferrites
are in phase with the source fields and the term is simply  $
\textbf{A}^2 $.  This a  consideration for conductors alone.

\bigskip

 With vector potentials in the
range of $ 10^{-5} $ to $ 10^{-10} $ tesla meter  for the range of
experiments the strength of the transitions depends very much on the
magnitude of \textbf{A}.

\bigskip

\textbf{Table 7:  Comparison of transition strengths, $m_e$ electron
mass and c is the speed of light in vacuum, set q for spin wave = 1.
The velocity of light in the metal is that of a transverse
electromagnetic field at $ ~ $ 10 kHz. This velocity is closer to
the speed of sound that the speed of light in vacuum. }
\begin{center}
  \begin{tabular}{|c|c|c|c|c| }
  \hline
\textbf{mass } &   \textbf{velocity} &  \textbf{q} & $\frac{2 e \hbar\textbf{q}}{ m c} \textbf{A}  $ &   $ \frac{e^2}{mc^2} \textbf{A}^2 $ \\
\hline
 $ m_e $ &  c  & 1 & $10^{-31} \textbf{A}$ & $ 10^{-25} \textbf{A}^2$\\
  mass spin wave $ 10^{-9}m_e $ & c  & 1& $10^{-22} \textbf{A}$ &  $10^{-16} \textbf{A}^2 $\\
  mass spin wave $ 10^{-9}m_e $ & $ \sqrt{\frac{\omega}{\mu\sigma}}  \sim 10^{-5}c $
  &1
  & $10^{-17} \textbf{A}$ & $10^{-11} \textbf{A}^2$ \\

\hline
  \end{tabular}
\end{center}
\bigskip

The role the reduced effective mass of the spin wave plays in
allowing transition due to the vector potential to the first or
second power is large.  The further enhancement by considering the
reduce velocity of light within the metal to drive transitions from
the source current is also significant compared to the coupling to a
free electron.    The next feature that plays a large role in the
interaction of the BELC is that their field is slowly varying over
the bar and the effective vector potential for the same \textbf{B}
level is greater by the ratio of the source penetration depth to the
diameter of the rod.  This further enhancement by a factor of 10 to
100 for the BELC field to drive transitions  will promote
interactions between BELCs.  At field levels for the vector
potential at $ 10^{-5} $ Tesla-meter  the linear and quadratic terms
in the Hamiltonian  approach each other in value.  This would then
be expected to allow the two types of transition of the kind found
when energy, linear and angular momentum are conserved.

 Now we can consider the transitions that are allowed
between the spin waves and the applied field solely. This is
summarized in the following table. The conditions are that linear
momentum and angular momentum must be conserved along with energy
and the phase of the interactions are assumed to be preserved. The
first transition to consider is that from the ground state. The
controlling element is the small linear momentum associated with a
photon $ \frac{\hbar\omega}{c} $ compared to that of the propagating
spin wave mode.

\bigskip

\textbf{Table 8:   Transition for the Linear Geometry}
\begin{center}
  \begin{tabular}{|c|c|c| }
  \hline
\textbf{Transition } &   \textbf{final state} & $ \Delta\textbf{J} $\\
\hline
$ < +q,\omega;-q,\omega | \textbf{A}(\omega)^2|0 > $  &  $ |0> \Rightarrow \psi_- $ ;    allowed, direct to a paired state  & 0    \\
$ < q,  \omega | \textbf{A}(\omega) | 0> $ &     \textbf{p}  not conserved  without phonon or other, not a pair process & 1    \\
$ <\omega_1 \pm  \omega_2 | \textbf{A}(\omega_2) | \omega_1 > $ &     \textbf{p} not conserved without phonon or other, not a pair process & 1   \\
$ <\omega_1 \pm  2 \omega_2 | \textbf{A}(\omega_2)^2| \omega_1 > $ &     \textbf{p} not conserved without phonon or other, not a pair process  & 1     \\
$ <\omega_1 \pm  \omega_2 | V_3| \omega_1; \omega_2  > $  &    allowed as a pair-pair process  & 0   \\
$ <\omega_1 \pm  2 \omega_2 | V_2| \omega_1; \omega_2 > $ &    two step pair-pair process          & 0       \\
$ 2 \psi_-  \Leftrightarrow   \overrightarrow{\psi}_+ +
\overleftarrow{\psi}_+ $ &  exchange of components &  0 \\

\hline
  \end{tabular}
\end{center}
\bigskip

 Transition to and from the vacuum state to
an excited spin wave state are allowed for all ferromagnetic
materials without considering phonon interactions and the scattering
of thermal spin waves into a BELC.  The last two transitions are
between BELC scattering interaction. These four transitions have
only been easily seen in well annealed iron based alloys or glasses
and were not evident in nickel or cobalt under similar test
conditions. Though the cobalt and nickel samples were not as well
annealed as the iron specimens. These four transitions look to be
from the scattering of BELC alone and not a direct interaction with
the applied field. The last two transitions require a spin wave in
both bands so that angular momentum is conserved. This would not be
possible in either nickel or cobalt and are not seen.

The field dependent matrix elements do not seem to operate over the
entire material to produce these paired states.  They seemed to be
limited to magnetic domain boundaries that can be perturbed by the
fields.  The surface of iron would not be a source of these paired
spin waves because it is not a region in which the local
magnetization goes through zero.  To apply a surface to iron that
would have this characteristic would make an enhanced source of
these waves and the sample would be more sensitive to coupling with
external fields without having the losses associated for a
penetrating field. If a material surface is coated with a thin anti
ferromagnetic layer over the iron, it maybe possible to use this new
interface with the iron as an enhanced source of these paired spin
waves.

\section{Conclusion}

With the application of a transverse time dependent induction,
$\textbf{H}(\omega)$, in iron and steel lifts the degeneracy in the
phase of the  magnetization   of the spin wave and allows spin pairs
across bands to couple forming $J=0$ states. The reduction of a
coupled pair to a $J=0$ states within frequency ranges removes the
strongest scattering interactions for the spin wave and allow its
wave length to scale upwards by five order of magnitude. This is
first detected by noting the effective mass of the spin wave has
been scaled back by ten orders of magnitude. This leaves this new
group of spin waves which are subject to Bose-Einstein statistics at
sufficient but easily achievable densities to allow a formation of a
Bose-Einstein like condensation all the way to the Curie point.

Two other time dependent magnetization dispersion curves were also
identified one with a coupled magneto-acoustic coupling that
operates when these long wavelength spin waves have velocities below
the velocity of sound and second dispersion curve found was a
measure of the density of a new state that had a net zero momentum.
This process of forming the BELC for the freely propagating spin
waves reduced the strong spin wave coupling to lattice as whole and
allows the formation of a collective zero momentum bound state. This
new field $ \psi_- $ show a strong response at high frequencies and
is also macroscopically large relative to the sample size.  From the
spatial representation of the BELC wave functions the inter particle
interaction energies were computed for these two spin wave fields
and are consistent from what would be expected from a stable BEC.
The BELCs formed are very weakly interacting with their environment
though they interact strongly with each other. Transition between
BELC states of different energy  enables  the creation of new
states.

The passage of these propagating magnetization through regions held
above the Curie point is not simply explained by one field or
another.  Though it does appear that $ \psi_-$ preferentially
traverse this region due to the phase drop. The fraction of signal
in each state as a function of temperature also has to known along
with  the gain   as a function of temperature.

Returning to the original problem of putting Maxwell's equations
into a form to handle these problems now looks to be tractable with
the solutions of the Gross-Pitaevskii  equations and model for BELC
populations dynamics from the coupling of the field into these
states and at low frequencies coupling in the elastic properties of
the material.  Notation is simplified by this approach to the
dynamic magnetic problem.   These ideas should be useful in studying
other ferromagnetic materials other than iron, which may not exhibit
as many features as were found here. These effects  will go along
way in explaining a number of measurements made in steels over the
years that were not understood. But of more general interest, the
mechanics of multiple BELC activity and boson mass are topics of
interest in other branches of physics.  The system studied here
after careful analytical development may yield more insights because
of the relative ease with which experiments can be done and ideas
tested.

\section{ \textbf{Acknowledgements} }

I would like to acknowledge the careful and extensive measurement of
high temperature eddy current responses in a variety of steels by
Mike Bergerhouse that firmly established this problem,  the
discussions of signal transmission and noise through steels taken
above the Curie point with Michael Wallace that gave clues  to the
character of the radiation source, Julian Nobel for giving me an
appreciations for the concept of mass and to Steven Wallace whose
patient methods of attack on problems was the example I followed.
Editing to reduce these data to a readable form was generously
performed by Jerry Dunn and Michael Wallace.

\bigskip
\bigskip

\appendix{ \textbf{Appendix A:  Single Homogenous Reflector} }

\bigskip

The simple case of a reflector that is a magnetic conductor or
insulator can be represented by a boundary value problem where the
electric field and the magnetic induction are required to be
continuous across the interface.  The vector potential of the source
field in free space will be represented by $ f $ the reflected
response will be $ g $ and the field propagating into the magnetic
medium is $ F $. The time dependence of the fields is  sinusoidal, $
e^{-i\omega t} $ as driven by the source.   The coefficient for the
field are \textbf{a}, \textbf{b} and \textbf{c} to produce the three
vector potentials $ \textbf{a}f $, $ \textbf{b}g $ and $ \textbf{c}F
$. The electric field  $ \textbf{E} = -i \omega \textbf{A} $ and the
magnetic induction $ H = \frac{\nabla\times \textbf{A}}{\mu}$.  In
this one dimensional representation the curl of the particular
vector potential will be represented by $ f' $, $ g' $ and $ F' $.
Dividing out the time dependence the vector potential one has the
relationship for the continuity for the transverse electric field:

$$ \textbf{a}f +  \textbf{b}g = \textbf{c}F    \eqno[1a]  $$

similarly for the continuity of the magnetic induction:

$$ \frac{\textbf{a}f' +\textbf{b}g'}{\mu_o}  =
\frac{\textbf{c}}{\mu}F   \eqno[2a] $$

We measure

 $$  V   =  \int \textbf{E} \cdot d\textbf{l}  = -i \omega\int
 \textbf{A} d\textbf{l}   \eqno[3a] $$

In the normalized form of our calibration the measured reflected
field  is

$$   V_{normalized }  =   -i \omega \int  \frac{ \textbf{b}g}{\textbf{a}f}
d\textbf{l}  = -i\omega\Lambda \frac{ \textbf{b}g}{\textbf{a}f}
\eqno[4a]  $$

where $ \Lambda $ is a length Solving the continuity equations to
eliminate \textbf{c} and define  \textbf{b} in terms of \textbf{a}.

$$ \textbf{a} \{ f - \frac{\mu F}{\mu_o F'}f' \}   =   \textbf{b} \{
\frac{\mu F}{\mu_o F}g' - g \}  \eqno[5a]  $$

For simplicity taking the case of an infinite planar reflector at $
x =  0 $  with $  f = e^{ikx } $,   $ g = e^{-ikx } $ and $ F =
e^{iKx } $ the above relation reduces to:

$$  \frac{\textbf{ b}}{\textbf{ a}} =   \frac{ \mu k - \mu_o K}{\mu k + \mu_o K}   \eqno[6a] $$

Taking in free space $ k =  \omega \sqrt{\epsilon \mu_o} $  and $ K
= \frac{i + 1 }{\sqrt{2}}\sqrt{\mu \sigma\omega} $ for the
conducting magnetic medium.  The result simplifies  where $ d $ is
just a constant.

$$ \frac{\textbf{ b}}{\textbf{ a}} =   \frac{\mu - d \mu^{.5}}{\mu + d \mu^{.5} }  =   \frac{\mu^{.5} - d }{\mu^{.5} + d
 }   \eqno[7a] $$

 Then in the limit of $  \mu \rightarrow \infty $ ,
 $\frac{\textbf{b}}{\textbf{a}}  \rightarrow 1 $.  This is true for
 magnetic conductor or insulator and also in the cylindrical
 geometry.

As a comment the two dimensional solution of a coil as a loop source
surrounding a cylindrical bar taken from reference 12 equation 63
can be computed for an empty coil and one filled with one
homogeneous materials.   The ratio of these two integral equations
can be taken in the limit of $  \mu \rightarrow \infty $ to show the
reflections are bounded.  If the same is tried for the loop above a
plane, the probe field, there is a miss print in equation 22 and 24
where the terms have not been divide by the the material
permeability so that before using equation 41 this derivation has to
be corrected.

\bigskip

\bigskip

\section{Bibliography}

1 H. Rowland, Phil. Mag., \textbf{48} p. 321 (1874).

2 E.M. Terry Phys.Rev. \textbf{30} no. 2 p. 133  (1910).

3 R.M. Borzoth, J. Appl. Phys,\textbf{ 8} p. 575  (1937).

4 H. J. Williams, R. M. Bozorth, W. Shockley,  Phys. Rev.
\textbf{75} p. 155 (1949).

5 J. Sola, US Patent, \textbf{2,694,177} Nov. 9 1954.

6 F. Brailsford, \underline{Physical Principles of Magnetism}, van
Nostrand, (London) 1966.

7 T.R. Schmidt, Mater. Eval. \textbf{42}, no 2 p. 225 (1984).

8 F Block, Z.Physik \textbf{61}, p. 206 (1930); \textbf{74} p. 295
(1932).

9 A.Sommerfeld, H. Bethe  \underline{Elektronentheorie der Metalle},
Springer, Berlin   1933.

10 W.B. Pearson,\underline{A Handbook of Lattice Spacings and
Structures of Metals and Alloys}, Vol. 1, p. 908. Oxford: Pergamon
Press 1967.

11 V.L. Moruzzi, J.F. Janak, A.R. Williams, \underline{Calculated
Electronic Properties of Metals} Pergamon Press p. 170 NYC 1978.

12 C.V. Dodd \& W. E. Deeds J. Appl. Phys. \textbf{39},2829 (1968).

13 L.D. Landau \& E.M. Liftshitz, \underline{Electrodynamics of
Continuous Media}, Trans J.B. Sykes, J.S. Bell p. 120  Pergamon
Press, Bristol 1960.

14 R. M. Siegfried, The Reconstruction of Electrical Conductivity
Profiles Using Multifrequency Eddy Current Testing, Thesis,
University of Minnesota, Mnpl. Minn. 1983.

15 J.P. Wallace, J.K. Tien, J.A. Steffani, K.S. Choe J. Appl. Phys.
\textbf{69} p. 550 (1991).

16 J.P. Wallace, US Patent, \textbf{4,651,094}  March 17, 1987, col
12-14.

17 M.E. Armacanqui, Eddy Current Detection of Sensitization in Types
304 and 316  Stainless Steels. Masters Thesis Univ. Minn. 1981.

18 C.Iheagwara, A Study of Transformation Kinetics in Cast Iron and
Slag using Eddy Current Technique, Masters Thesis Univ. Minn. 1982.

19 Mike Bergerhouse, private communications 1995.

20 G. Shirane, R. Nathens, O. Steinsvoll, H.A. Alperin, S.J.
Pickart, Phys. Rev. Lett. \textbf{15} p 146 (1965).

21 A.I. Akhiezer, V.G. Bar'yakhtar,S.V. Peletminskii,
\underline{Spin Waves} trans S. Chomet  trans. ed. S. Doniach, p.
134, North-Holland Pub., Amsterdam (1968).

22 H. Bethe, Z. Physik \textbf{61} p. 205 (1931).

23 S.O.Demokritov, V.E. Demidov, O. Dzyapko, G.A. Melkov, A.A.
Serga, B. Hillebrands and A.N. Slavin Nature \textbf{443}, p. 430
(2006).

24 L.D. Landau, E. M. Lifshitz, \underline{Statistical Physics},
trans E. Peierls and R.F. Peierls, Pergamon Press, London 1958.

25 F. Dalfovo, S. Giorgini, M. Guilleumas, L. P. Pitaevskii, S.
Stringari   Rev.Mod.Phys. \textbf{71}, p. 463 (1999).

26 A.A. Thiele Phys. Rev. B. \textbf{7} p. 391 (1973).

27 J.P. Wallace, Journal of Metals,\textbf{61} no. 6 p. 67 (2009).

28 K. Honda, S.Kaya, Sci.Rep. Tohoku Univ. \textbf{15} p. 721
(1926).

29 J. Schwinger \underline{Quantum Mechanics}, ed. B-G. Englert, p.
321, Springer, Berlin 1965.
\medskip

\end{document}